\documentclass[12pt]{biophys}
\usepackage[utf8]{inputenc}
\usepackage{bm,bbm,graphicx,epstopdf,setspace}
\usepackage{fancyhdr}
\usepackage{graphicx,subfigure,wrapfig,color,enumerate}
\usepackage{amsmath, amssymb,amsfonts,psfrag,latexsym}
\usepackage{siunitx}
\usepackage{cancel,soul}
\usepackage[numbers,square,sort&compress]{natbib}
\usepackage{url,fancybox}
\usepackage[linktocpage=true]{hyperref}
\hypersetup{colorlinks,citecolor=blue,linkcolor=red,urlcolor=blue}
\usepackage{lipsum}
\usepackage{helvet,times}
\usepackage{bm,textcomp}
\jno{kxl014} 
\gridframe{N}
\cropmark{N}
\doi{doi: 10.1529/biophysj.xxx.xxxxxx}

\def\str{SB}
\def\inter{IB}
\def\weak{WB}
\def \kbt{$k_{\rm B}T$}
\def \kbtt{k_{\rm B}T}
\def \kb{k_{\rm B}\,}

\def\kbt{$k_{\rm B} T$}
\def\ekbt{\,k_{\rm B} T}
\def\molnc{${\cal W}_{\rm NC}$}
\def\molncc{{\cal W}_{\rm NC}}
\def\molpmf{${\cal H}_{\rm SMD}$}
\def\avm{$\langle m \rangle$}
\def\eavm{\langle m \rangle}

\newcommand{\eNab}[2][]{$N_{\rm ab}$ /NC}

\newcommand{\Nantt}[1][]{$N_{\rm ant}$}
\newcommand{\dr}[1][]{$\Delta R$}
\newcommand{\drr}[1][]{\Delta R}
\newcommand{\sqmic}[1][]{$\mu {\rm m}^{2}$}
\newcommand{\sqmicc}[1][]{\mu {\rm m}^{2}}

\newcommand{\nab}{N_{\rm ab}}
\newcommand{\nant}{N_{\rm ant}}
\newcommand{\astrm}{\si{\angstrom\,}}
\newcommand{\sinfo}[1]{{\color{blue} Suppl. Info., Sec. #1}}

\begin{document}

\setcounter{page}{1} 

\title{Computational investigation of multivalent binding of a ligand coated particle: Role of shape, size and ligand heterogeneity from a free energy landscape perspective}

\author{Matt McKenzie$^{1}$, Sung Min Ha$^{2}$, Aravind Rammohan$^{1}$,  Ravi Radhakrishnan$^{2,3,4}$, \& N. Ramakrishnan$^{2,*}$}
\address{ 
	$^{1}$ Corning Inc. Corning, NY, USA
	$^{2}$ Department of Bioengineering, University of Pennsylvania, Philadelphia, PA, 19104, USA, 
	$^{3}$ Department of Chemical and Biomolecular engineering, University of Pennsylvania, Philadelphia, PA, 19104, USA, 
	$^{4}$ Department of Biochemistry and Biophysics, University of Pennsylvania, Philadelphia, PA, 19104, USA \\
	\vspace*{10pt}
	$^{*}$ corresponding author: ramn@seas.upenn.edu
}

\begin{abstract}
{We utilize a multiscale modeling framework to study the effect of shape, size and ligand composition on the efficacy of binding of a ligand-coated-particle to a substrate functionalized with the target receptors.  First, we show how molecular dynamics (MD) along with Steered MD calculations can be used to accurately parameterize the molecular binding free energy and the effective spring constant for a receptor-ligand pair. We demonstrate this  for two ligands that bind to the $\alpha_5\beta_1$-domain of integrin. Next, we show how these effective potentials can be used to build computational models at the meso- and continuum- scales. These models incorporate  the molecular nature of the receptor-ligand interactions and yet provide an inexpensive route to study the multivalent interaction of  receptors and ligands through the construction of Bell potentials customized to the molecular identities. We quantify the binding efficacy of the ligand-coated-particle in terms of its multivalency, binding free energy landscape and the losses in the configurational entropies. We show that (i) the binding avidity for particle sizes less than $350$ nm is set by the competition between the enthalpic and entropic contributions while that for sizes above $350$ nm is dominated by the enthalpy of binding, (ii) anisotropic particles display higher multivalent binding compared to spherical particles and (iii) variations in ligand composition can alter binding avidity without altering the average multivalency. The methods and results presented here have wide applications in the rational design of functionalized carriers and also in understanding cell adhesion.}
	
\end{abstract}

\maketitle


\section{Introduction}
Specific receptor-ligand interactions is a recurring theme in cell biology. A cell utilizes this process to respond to external cues by switching on or off signaling cascades and gene expression programs, to govern and mediate its interactions with the cytoskeleton, the extracellular matrix, and also with other cells and to commit to different cell fates. Receptor-ligand interactions are known to be one of the key effectors of receptor trafficking in cells and are also used by external particles such as a virus to gain its entry into the cytoplasm~\cite{Dimitrov:2004cx}. The principle of selective interactions between a pair of receptor and ligand has also been exploited in technologies such as targeted medicine and contrast imaging using functionalized nanoparticles (NPs) or  nanocarriers (NCs) that are used as vehicles for  delivery of therapeutic agents or for selective localization of diagnostic agents~\cite{Peer:2007kb,Mitragotri:2014by,Shi:2016fg}.

{Specific receptor-ligand adhesion has been extensively studied in the literature across various spatial and temporal scales~\cite{Hynes:1994tz,Hynes:1999wx,Rangarajan:2008jv,Khalili:2015kq}. At the cellular scale (characteristic length $>$ $1$ $\mu$m), the formation of focal adhesions due to the interaction of cell-surface integrins with the extra-cellular-matrix or to ligand patterned on  substrates~\cite{Hynes:1992tg,Garcia:1998wx,Clark:2005fj,RocaCusachs:2009hv,Nieto:2011db,RocaCusachs:2012dl,Bendas:2012di,Coyer:2012gv,RocaCusachs:2013hq,Schaffner:2013jn,Bharadwaj:2017jk} has been a topic of major interest for experimental investigations of cellular adhesion. These studies have been well complemented with theoretical and computational models at the cellular scale~\cite{Xiao:1996kb,Irvine:2002ix,Mallet:2006co,Atilgan:2009hs,Cirit:2010ib,Wong:2010gx,Care:2011ji,Welf:2012fs,Peng:2012kz} that have helped to gain deeper insight into the various physico-chemical interactions governing cellular adhesion and its role in motility, signaling, metabolism and mechanotransduction. At the mesoscale (characteristic length $\sim$ $100$ nm) , experimental studies have largely focused on the biodistribution of functionalized NPs, targeted to various cellular adhesion molecules both \textit{in vitro} and \textit{in vivo}, for use as targeted drug delivery systems~\cite{Agarwal:2013bza, Lesniak:2013em,Agarwal:2013bz,Kolhar:2013kd,Agarwal:2015fs,Chaudhary:2015bl,Cox:2016ks,Pitchaimani:2017cd}. A number of theoretical and computational models~\cite{Weikl:2009kt,Liu:2010em,Shah:2011hga,MartinezVeracoechea:2011kn,Liu:2012ve,Liu:2012wn,Bahrami:2013km,Dasgupta:2014hr,AgudoCanalejo:2015gy,Dubacheva:2015hc,Tito:2016dk,Ramakrishnan:2016fl,Yoon:2016fa,Schubertova:2017bz} have been developed to address the question of adhesion and super-selective targeting of functionalized NCs and their uptake into the target cell~\cite{Vacha:2011cf,Vacha:2012bd,Huang:2013eh,GonzalezRodriguez:2015ek,Schubertova:2015vb,Guo:2016iz}. At the molecular scale (characteristic length $<$ $20$ nm), all atom molecular dynamics simulations have been extensively used to unravel the molecular mechanisms governing receptor-ligand interactions~\cite{Gao:2004bw,Craig:2004kp,Krammer:1999cy,Nagae:2012bz,Kalli:2016fk} and the interaction of gold and silver nanoparticles with the surface of a cell~\cite{VanLehn:2013ky,VanLehn:2014hb,Monticelli:2016bi,Rossi:2016cg}.}

It has been well established that the adhesion of functionalized particles with a substrate strongly primarily depend on their shape, size and surface  chemistry~\cite{Blanco:2015ek, Moghimi:2012hy,Nel:2009ie,Shang:2014bp}. The latter is primarily determined by the specific interactions between the particle surface and the receptors on the substrate, and to a very small extent is also determined by the non-specific interactions between the particle and substrate. The ``substrate'' here can refer to a cell surface in culture, live cells in a tissue, or cells on enhanced tissue culture scaffolds such as matrices, flasks and stacks~\cite{Griffith:2006df,Eglen:2015dk}. {In our earlier studies, we have shown that the binding affinity of a functionalized NP  is jointly determined by the competition between the binding enthalpy (that enhances binding) and the loss in configurational entropy of the various components (that inhibits binding)~\cite{Liu:2010em,Ramakrishnan:2016dk}. This feature is particularly pronounced for NCs with characteristic size in the mesoscale. The primary objective of this article is to develop a comprehensive picture of this phenomenon and delineate the enthalpy dominated  regimes from the entropy dominated  regimes for particle binding.}


One of the key challenges in developing such a picture is to address the multiscale nature of the various processes that promote particle adhesion. For example, a spherical particle of radius $50$ nm functionalized with $\sim 150$ ligands whose radius is $\sim 3$ nm  is typically used in targeted experiments~\cite{Liu:2012wn,Muzykantov:2013jg} for which the surface ligand density is around $14\%$. This is typical of biomedical applications due to (i) the large cost of antibodies and (ii) the risk of triggering immune response at higher densities. At such low ligand densities it is only appropriate to treat the interactions between the ligand on the particle and receptors on the target substrate as patchy, rather than as a continuum. This would  require a different class of models in which we retain the molecular nature of the receptor-ligand interactions. In the past, researchers have treated NC adhesion either as a continuum~\cite{Weikl:2009kt,Bahrami:2013km,Dasgupta:2014hr} or as semi-continuum~\cite{Liu:2010em,Ramakrishnan:2016fl} to quantify the specific adhesion of functionalized particles with rigid and flexible substrates. A main ingredient, namely entropy loss of receptors facilitating multivalent binding, is absent in such continuum models, which we address using our approach. Moreover, in all of these works, the framework for incorporating the molecular identities of the receptor-ligand interactions was primarily determined from experiments~\cite{Liu:2010em}. However, as shown in this work, a computational route to include the molecular identities can also be achieved using the molecular dynamics framework.

\begin{figure}[!h]
\centering
\includegraphics[width=7.5cm,clip]{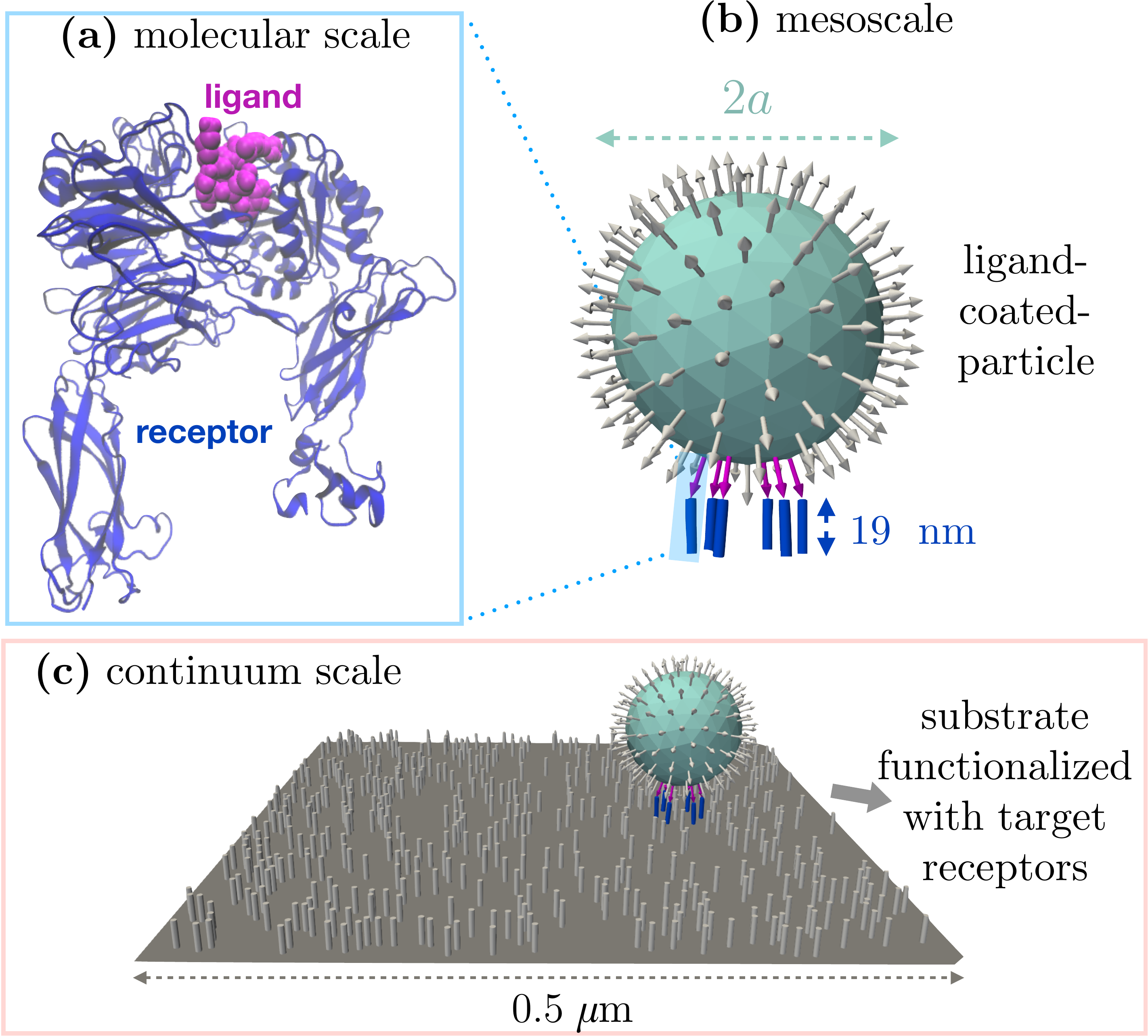}
\caption{\label{fig:overview} Multiscale aspects of modeling ligand coated particles binding to a substrate functionalized with the target receptors. (a) A snapshot of a receptor-ligand interaction at the molecular scale, (b) multivalent binding of a ligand coated spherical particle -- the ligand (arrows) and receptors (cylinders) that mediate the binding are colored magenta and blue, respectively, and (c) a snapshot of our continuum scale simulations to estimate binding efficacy.}
\end{figure}

In this article, we {utilize} a generalizable multiscale computational framework to study the multivalent adhesion of ligands immobilized on a particle surface with receptors diffusing on a substrate, as displayed in Fig.~\ref{fig:overview}. Our multiscale approach consists of two steps: (i) molecular scale characterization of monovalent receptor-ligand interactions (Fig.~\ref{fig:overview}(a)), and (ii) a mesoscale model for multivalent receptor-ligand interactions (Fig.~\ref{fig:overview}(b)). Details of the multiscale coupling between the molecular scale and the mesoscale are described in the methods section. We systematically investigate the role of (i) receptor-ligand interaction strength, (ii) particle shape, (iii) particle size and (iv) ligand concentration  on the binding avidity of the NC. In this work, we will only focus on NC adhesion to rigid substrates, while its extension to flexible substrates is straightforward as shown in our earlier work~\cite{Ramakrishnan:2016fl}. However, here, we do provide a computationally efficient approximation to consider the effect of substrate compliance, such as when the NC binds a deformable cell membrane in a live cell by building the effect of compliance into the interacting potential. We quantify the equilibrium bound state of a NC bound to a substrate in terms of (i) the losses in the  configurational entropies and (ii) the gain in binding enthalpy. We explicitly compute these terms and delineate the entropy-dominated and enthalpy-dominated regimes for NC binding.
\section{Receptor-ligand pairs of interest}
We will focus on three specific receptor-ligands pairs, denoted as (i) strong binder (\str), (ii) intermediate binder (\inter), and (iii) weak binder (\weak). The \str{} system corresponds to an Inter-cellular adhesion molecule (ICAM aka CD54) receptor interacting with its YN1 antibody, while the \inter{} and \weak{} systems correspond to the $\alpha_5\beta_1$ domain of integrin interacting with peptide sequences KGGEP\underline{RGD}TYR and GE\underline{RGD}GSFFAFRSPF, both of which contain the well-recognized RGD (Arg-Gly-Asp) binding domain. These sequences are peptidomimetics that mimic the integrin binding domains found in the extracellular matrix proteins, fibronectin and collagen, respectively~\cite{Kotamraj:2011kr,Healy:1995vx,Hersel:2003to,Melkoumian:2010if}. The interaction energies of each of these pairs are listed in Table~\ref{table:system}, and the snapshots of the IB and WB systems in their respective bound states are shown in \sinfo{S1}.

\begin{table*}[!t]
\centering
\footnotesize{ 
\begin{tabular}{c|c|c|ccc}
\hline
&  & & \multicolumn{3}{c}{Binding energy, $\Delta {\cal H}_b$ (\kbt)} \\
System & Receptor & Ligand sequence &  \textit{EXPERIMENTS} & \textit{AUTODOCK} & \textit{SMD}  \\
\hline
\hline
Strong binder (\str)& ICAM-1  & anti-ICAM1 (YN1)& $-19.1$~\cite{Liu:2010em} & - & -  \\
Intermediate binder (\inter)   & $\alpha_5\beta_1$ domain of Integrin & KGGEP\underline{RGD}TYR& $-11.45$~\cite{Kotamraj:2011kr} & $-12.45$ & $-12.22$   \\
Weak binder (\weak) & $\alpha_5\beta_1$ domain of Integrin & GE\underline{RGD}GSFFAFRSPF& $-1.76$ ~\cite{Healy:1995vx} & $-1.82$ & $-1.71$ 
\end{tabular}
}
\caption{\label{table:system} \str, \inter, and \weak, the three receptor-ligand systems used in our study, along with the corresponding binding energies (in \kbt{}). The experimental values have been taken from the literature while the computational values are our estimates obtained using Autodock and steered molecular dynamics (SMD) simulations. The details of these calculations are given in the methods section, and discussions on the Autodock and SMD results may be found in the results sections.}
\end{table*}
Our choice of these pairs of receptors and ligands has been motivated by the fact that : (i) their interaction energies represent three different energy scales that spans over the physiologically relevant values for receptor-ligand interactions ($\sim -2$, $-10$ and $-20$   \kbt{} for \weak{}, \inter{} and \str{}, respectively, as shown in Table ~\ref{table:system}) and (ii) the wide applicability of these results both in the development of functionalized NCs and also in understanding cell adhesion with its substrate.  Since the \str{} system, that is comprised of the ICAM1-YN1 receptor-ligand pair, has been previously well characterized~\cite{Agrawal:2007bma,Liu:2010em,Liu:2011fe,Liu:2011bf,Ramakrishnan:2016fl}, we will demonstrate the applicability of our molecular characterization techniques only for the \inter{} and \weak{} systems. Another benefit of considering SB, IB and WB is that it enables us to account for the phenomenon of opsonization (deposition of albumin or FC receptor fragments from the plasma onto the NC surface)~\cite{Blanco:2015ek,Moghimi:2012hy,Lundqvist:2008ge}. In this case, while the functionalized molecule can be modeled as a SB, the opsonized moieties can be treated either as a IB or WB.
\section{Methods}
We use a bottom-up approach to self-consistently estimate the binding affinity of a functionalized NC. In this approach, we first characterize the monovalent interactions of a receptor-ligand pair using molecular dynamics techniques. The molecular scale interaction potentials are then used as an input for the mesoscale models to estimate the thermodynamic stability of the multivalent receptor-ligand bonds. In this section, we present the details of the various models and simulations techniques employed in each of these length scales.
\subsection{Molecular methods and atomistic potentials}
\noindent(i) \textit{Atomistic models for \inter{} and \weak{} systems}: The structure of the $\alpha_5\beta_1$ domain of integrin, which is the receptor for both the \inter{} and \weak{} systems, was extracted from its complex with RGD peptide obtained from the RSCB Protein Data Bank (\href{https://www.rcsb.org/pdb/explore.do?structureId=3VI4}{pdb id: 3VI4})~\cite{Nagae:2012bz}. The corresponding ligand sequences for the \inter{} and \weak{} systems were prepared using standard procedures. In our model for the receptor, we also retained the Ca$^{2+}$ and Mg$^{2+}$ ions  found to be co-crystallized in the 3VI4 structure. {In our simulations we use standard protocols defined under the AMBER force field to account for the difference between Ca$^{2+}$ and Mg$^{2+}$ ions. For non-conventional force fields care should be exercised to make this distinction.}

\noindent (ii) \textit{Docking studies}: The binding sites for the  \inter{} and \weak{} systems were determined using AutoDock~\cite{Morris:2009gb}, a genetic algorithm based conformer search tool. The AutoDock search was performed using a $30$ \astrm cubic sampling box, centered on the active site of the integrin headpiece, with a grid size of $0.375$ \astrm. The integrin was kept fixed during this procedure and the initial ligand population was set to $200$. For each system, the complex with the highest docking score was taken to be the bound structure of the receptor-ligand complex and these were used as the initial configuration for their respective molecular dynamics (MD) simulations. 

\noindent (iii) \textit{MD simulations and generation of equilibrium bound structures}: We performed molecular dynamics simulations of the bound complexes using GROMACS$4.6$~\cite{Pronk:2013ef} with the Amber$99$SB-ILDN protein-nucleic AMBER$94$ force field~\cite{LindorffLarsen:2010ei} for the organic molecules. The bound complexes were solvated in SPC/E water~\cite{Berendsen:1987uu} in a periodic cubic box of length $70$ \astrm. The total number of atoms inclusive of the complex, counterions, and water were over $400,000$. The equilibrium bound structure for both the \inter{} and \weak{} systems were obtained by (a) energy minimization in the NVT ensemble followed by (b) equilibration in the NPT ensemble at $300$ K for $10$ ns, with a $2$ fs timestep; here the hydrogens were constrained using the SHAKE algorithm. A representative snapshot of the equilibrium receptor-ligand conformation for the \weak{} system is shown in Fig.~\ref{fig:overview}(a).

\noindent (iv) \textit{Steered molecular dynamics (SMD) and Umbrella Sampling}: We performed SMD calculations on the equilibrium bound state structures of the receptor and ligand to compute a representative binding path for the \inter{} and \weak{} systems. In both cases, the distance between the center of masses of integrin receptor and its ligand was taken to be the reaction coordinate for the SMD calculations. We performed a 10 ns SMD in the NPT ensemble with a pulling rate of $0.01$ nm/ps and a pulling spring constant of $1000$ kJ/(mol nm$^2$).  We stored the snapshot of the system at every time step and these snapshots were later used as the starting configuration for the umbrella sampling. In an umbrella sampling simulation, we first chose 30 different snapshots (sampling windows), each separated by  a window size of $0.15$ \astrm and added appropriate center of mass constraints. Next, for each of the snapshots we performed a 2 ns NPT equilibrium run followed by a 10 ns NPT production run and collected statistics for the reaction coordinate. The histograms of the reaction coordinate in all the windows were converted to a potential of mean force using the Weighted Histogram Analysis method (WHAM)~\cite{Kumar:1992hu,Roux:1995vi}. The obtained potential of mean force was taken to be the atomistic potential for receptor-ligand binding.
\subsection{Mesoscale model for multivalent receptor-ligand interactions}
We adopt the coarse grained approach developed in earlier works~\cite{Agrawal:2007bma,Liu:2010em,Liu:2011fe,Ramakrishnan:2016fl} to study the interaction of a ligand-coated NC with cell surface receptors. A snapshot of our mesoscale model is shown in Figs.~\ref{fig:overview}(b) and (c). In our representation, the NC is modeled as a spheroid with three principle dimensions $a$, $b$ and $c$, with $b=c$. We define the particle aspect ratio as $\varepsilon=a/b$, such that $\varepsilon=1.0$ for spherical particles, $\varepsilon<1.0$ for oblate spheroids and $\varepsilon>1.0$ for prolate spheroids. The target substrate is taken to be a square, rigid planar surface, of length $L_s$. We coarse grain each receptor and ligand molecule as flexible rods of lengths $L_r$ and $L_l$, respectively; these are shown as cylinders and arrows in Fig.~\ref{fig:overview}(c). 

The $N_l$ ligands are uniformly distributed and immobilized on the surface of the NC  and are oriented to be along the local surface normal (see \sinfo{S2} for details). The $N_{r}$ receptor molecules are randomly distributed on the planar substrate and are constrained to diffuse only along the $xy$-plane; in general, the receptors diffuse on the two dimensional curvilinear manifold defined by the flexible cell membrane~\cite{Ramakrishnan:2016fl}.  The total flexural energy of the receptors is given by:
\begin{equation}
{\cal H}_{\rm flex} = \dfrac{1}{2} \sum_{i=1}^{N_{r}} {\cal K}_{\rm flex} \left(L_r \sin\theta_{i}\right)^{2}.
\label{eqn:Hflex}
\end{equation}
${\cal K}_{\rm flex}$  is the flexural stiffness of the receptors and $\theta_{i}$ is the flexure angle for the $i$ th receptor, measured with respect to the $z$ direction (taken to be the unflexed orientation of the receptors). Since the ligands are immobilized on the surface of the NC we ignore any contributions due to ligand flexure. In an earlier work, we showed that the flexural stiffness of the receptor can be calculated using MD simulations~\cite{Liu:2011bf}.

Each ligand on the particle surface interacts with its receptors on the substrate and can form a bond, as shown in Fig.~\ref{fig:overview}(b); the total number of bonds formed by the functionalized ligands is defined as the multivalency of the particle. The interaction between a receptor and a ligand is modeled as a Bell potential~\cite{Bell:1978vf,Bell:1984kp}:
\begin{equation}
{\cal H}_{\rm bond}= \left\{  
\begin{array}{cc}
0 & \textrm{when } d \geq d^{*} \\
\Delta {\cal H}_{b} + \dfrac{{\cal K}_{b}}{2} d^{2}  &  \textrm{when } d<d^{*}
\end{array}
\right..
\label{eqn:Hbond}
\end{equation}
Here, $d$ is the distance between the tips of the receptor and the ligand and $d^*=\sqrt{2\Delta {\cal H}_b/{\cal K}_b}$ is the cutoff distance above which the receptor and ligand are in an unbound state. $\Delta {\cal H}_b$ is the binding energy   and ${\cal K}_b$ is the stiffness of the receptor-ligand bond. The values of $\Delta {\cal H}_b$ and ${\cal K}_b$ for \inter{} and \weak{} systems are directly determined from the atomistic potentials constructed using molecular dynamics simulations, which enables a coupling between atomistic scale and mesoscale models.

The total energy of a NC bound to a substrate with $m$ multivalent bonds is then given by:
\begin{equation}
{\cal H} = \sum\limits_{i=1}^{m} \left( {\cal H}_{\rm flex}(\theta_i) +  {\cal H}_{\rm bond}(d_i) \right).  
\label{eqn:Htotal}
\end{equation}
\noindent\textbf{Configurational degrees of freedom}: The configurational space for the mesoscale adhesion model is comprised of:
\begin{enumerate}
\item The position of the center of mass of the NC, in Cartesian coordinates, ${\bf X}_{p} = \left\{X_{p}, Y_{p}, Z_{p} \right\}$.
\item The orientation of the NC, in terms of the Euler angles,  ${\bf \Theta}_{p} = \left\{\phi_p, \theta_{p}, \psi_{p} \right\}$.
\item The spatial position of all receptors, in Cartesian coordinates,  ${\bf X}_{r} = \left\{\left\{x_{1}, y_{1}\right\},\cdots,\left\{x_{N_{r}}, y_{N_{r}}\right\}  \right\}$.
\item The orientation of all receptor molecules, ${\bf \Theta}_{r} = \left\{\left\{\theta_{1}, \phi_{1}\right\},\cdots,\left\{\theta_{N_{r}}, \phi_{N_{r}}\right\}  \right\}$, in the spherical polar coordinates.
\item The bonds formed between the receptors and ligands $\left\{ b \right \}$.
\end{enumerate}
The $x$ and $y$ dimensions are subject to periodic boundary conditions while the boundary along the $z$ direction is taken to be a hard wall.

\noindent\textbf{Monte Carlo moves to evolve the various degrees of freedom}: We evolve the system in its configurational space through a set of five independent Monte Carlo moves. Each move is randomly chosen and is designed to alter only one degree of freedom while holding the others fixed. 
\begin{enumerate}
\item \textit{Diffusion of the NC}: The center of mass of the NC  is moved to a new random position ${\bf X}_{p}+\delta {\bf X}_{p}$ within a cube of size $\epsilon$, centered about ${\bf X}_{p}$. The move is accepted using the Metropolis scheme~\cite{Metropolis:1953in} according to the probability $\min\left(1,\exp(-\beta \Delta {\cal H}) \right)$  where  $\Delta {\cal H}$ is the change in the total energy (eqn.~\eqref{eqn:Htotal}) due to the change in position and $\beta=\left(\ekbt\right)^{-1}$. Here $\kb$ denotes the Boltzmann constant and $T$ the absolute temperature.
\item \textit{Rotation of the NC about its own axis}: The orientation of the NC is changed from $\left\{{{\phi}_{p},\theta}_{p},{\psi}_{p} \right\}$ to a random orientation $\left\{ {\phi}_{p}+\delta {\phi}_{p} , {\theta}_{p}+\delta {\theta}_{p},{\psi}_{p}+\delta {\psi}_{p}  \right\}$, subject to the constraint  $0<\phi_p+\delta {\phi}_{p}<2\pi$, $0<\theta_p + \delta {\theta}_{p}<\pi$, and $0<{\psi}_{p}+\delta {\psi}_{p}<2\pi$. The new orientation  is accepted according to the Metropolis acceptance scheme. For particles with aspect ratio $\varepsilon \neq 1$, we represent the rotational degrees of freedom in terms of quaternions~\cite{Allen:1989ut,Kuffner:2004fv} and evolve them using a hybrid molecular dynamics-Monte Carlo scheme, as described in \sinfo{S4}.
\item \textit{Receptor diffusion on the substrate}: The position of a randomly chosen ligand  is changed from ${\bf X}_{r}$ to ${\bf X}_{r}+\delta {\bf X}_{r}$, where the components of $\delta {\bf X}_{r}$ are randomly chosen within a square of size $\eta$ centered about ${\bf X}_{r}$. The move is accepted according to the Metropolis scheme.
\item \textit{Receptor flexure}: The orientation of a randomly chosen ligand $i$ is changed from $\left\{ \theta_{i}, \phi_{i}\right\} \rightarrow \left\{ \theta_{i}+\delta \theta_{i} , \phi_{i}+\delta \phi_{i} \right\}$, where the randomly chosen increments are such that $0<\theta_{i}+\delta \theta_{i}<\pi$  and  $0<\phi_{i}+\delta \phi_{i}<2\pi$. The move is accepted using the Metropolis scheme.
\item \textit{Formation and breakage of receptor-ligand bonds}:  A bond is formed between a randomly chosen receptor-ligand pair if they are previously unbound, and the bond is broken/retained (with equal probability) if the chosen pair is already in the bound state. This move is performed and accepted by a configurational bias Monte Carlo move using the Rosenbluth sampling technique~\cite{Frenkel:2001}. The  configurational bias scheme for a chosen receptor $i$ and ligand $j$ is implemented by generating $200$ trial orientations for the receptor (where  $\theta$ and $\phi$ are chosen randomly) and $200$ random trial bound states ($1$ for bound and $0$ for unbound). The selected trial orientation is then accepted using the Metropolis scheme for configurational bias~\cite{Frenkel:2001}.
\end{enumerate}
The parameters $\epsilon$  and $\eta$  are chosen and adjusted during runtime so that half of the attempted moves are accepted.

\noindent \textbf{Mesoscale interaction potentials}:
We use umbrella sampling and WHAM to compute the potential of mean force (PMF) for a NC bound to a planar substrate. In our calculations, we take $z$-separation between the planar substrate and the center of mass of the NC as the reaction coordinate. If the planar substrate is taken to be at $z=0$, the reaction coordinate for the PMF calculations is given  by $Z_{p}$.  The umbrella sampling calculations are performed with an additional biasing potential:
\begin{equation}
{\cal H}_{\rm bias} = \dfrac{{\cal K}_{\rm bias}}{2} \left(Z_{p}-Z_{p,0} \right)^{2}.
\label{eqn:Hbias}
\end{equation}

Here $Z_{p,0}$ is the  position of the window for which the umbrella sampling is performed and the PMF is self consistently determined from the histograms of $Z_{p}$ using WHAM~\cite{Kumar:1992hu,Roux:1995vi}. In our calculations, the PMF is a direct measure of the absolute free energy for binding since the unbound states (where ${\cal H}=0$) and the bound states (where ${\cal H}\neq0$) are well defined along our choice of the reaction coordinate.

All our studies  were performed with a constant ligand density of $14\%$ and a constant receptor density of $2000$ receptors/$\mu{\rm m}^2$, as reported in experiments~\cite{Muro:2006fu,Muzykantov:2012uy}. The dimension of the substrate $L_s$ was taken to be five times larger than the dimension of the NC (i.e., $L_s=5a$ for $\varepsilon \ge 1$ and $L_s=5b$ for $\varepsilon<1$). A typical equilibrium simulation was run for a total of $10^9$ steps which were distributed between moves $1$ to $5$ at a ratio of $10\%$, $10\%$, $10\%$, $20\%$ and $50\%$, respectively. The free energy simulations were run for $3\times10^8$ steps per window with a similar distribution for the different classes of moves. All simulations were run in quadrupulates for realizing different ensembles and computing statistical errors. The errorbars in the free energy calculations denote the standard deviation over the four ensembles. The typical run time for an equilibrium simulation on a $2.6$ GHz processor is about $24$ hours, while that for each window in the free energy calculations is $10$ hrs. 
\section{Results}
\subsection{Characterization of receptor-ligand interactions and parametrization of coarse grained potentials }

\begin{figure}[!h]
	\centering
	\includegraphics[width=7.5cm,clip]{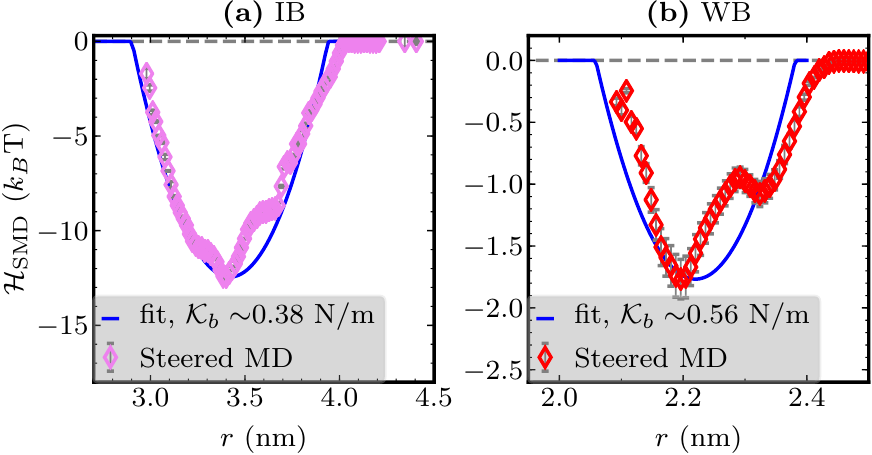}
	\caption{\label{fig:molecular-pmf} The molecular potential of mean force (\molpmf{}) for (a) the IB  and  (b) the WB systems. The solid lines denote the best fit of the data to a Bell bond potential, see eqn.~\eqref{eqn:Hbond}. The values of ${\cal K}_b$ for the IB and WB systems are found to be $0.38$ N/m and $0.56$ N/m, respectively.
	}
\end{figure}

\noindent\textbf{Computed values of receptor-ligand binding energies compare well with experimental estimates:} We computed the binding free energies ($\Delta{\cal H}_b$) for the \inter{} and \weak{} systems using two different approaches: (i) docking studies and (ii) SMD calculations, as described in the methods section.  It is straightforward to compute $\Delta {\cal H}_b$ using docking studies: we found $\Delta {\cal H}_b=-12.45$ and $-1.82$ \kbt{} for the \inter{} and \weak{} systems, respectively; while the corresponding values from the SMD calculations were determined as follows. ${\cal H}_{\rm SMD}$, the potential of mean force (PMF) for single receptor-ligand interactions computed using SMD calculations,  for both the \inter{} and \weak{} systems are displayed in Figs.~\ref{fig:molecular-pmf}(a) and (b), respectively. We take the binding energy to be equal to the minimum value of the PMF, i.e., $\Delta {\cal H}_b= \min({\cal H}_{\rm SMD})$. From the PMFs in Fig.~\ref{fig:molecular-pmf} we estimated $\Delta {\cal H}_b=-12.22$ \kbt{} for \inter{} and  $\Delta {\cal H}_b=-1.71$ \kbt{} for \weak{}. The values of $\Delta {\cal H}_b$ computed through the two methods are in excellent agreement with each other and also with experimentally obtained binding energies~\cite{Kotamraj:2011kr,Healy:1995vx}; this comparison is displayed in Table~\ref{table:system}.
 
\noindent\textbf{Explicit calculations show that the receptor-ligand interactions follow a Bell potential:}  The PMFs shown in Fig.~\ref{fig:molecular-pmf}, as pointed out earlier, represent the free energy landscape for monovalent receptor-ligand binding. The computed values of ${\cal H}_{\rm SMD}$ may then be directly used in place of eqn.~\eqref{eqn:Hbond} to model the receptor-ligand interactions in our mesoscale model for NC adhesion. In this approach, ${\cal H}_{\rm SMD}$ should be resolved at a very high spatial accuracy which leads to increased computational cost. As an alternative, we approximate ${\cal H}_{\rm SMD}$ to an analytic function which allows for an equivalent but efficient representation of the receptor-ligand interactions. Upon closer observation it may be seen that the PMFs for both the \inter{} and \weak{} systems resemble the Bell potential~\cite{Bell:1978vf,Bell:1984kp}, given by eqn.~\eqref{eqn:Hbond}. Others have shown experimentally that the Bell potential is a good representation of SB interactions~\cite{Hanley:2003iz,Florin:1994el,Liu:2010em}. Using the values of $\Delta {\cal H}_b$ determined earlier, we fitted  eqn.~\eqref{eqn:Hbond} to the molecular PMFs  and the best fit curves for the two systems are shown as solid lines in Fig.~\ref{fig:molecular-pmf}. From the fitting procedure, we determined the spring constants for the \inter{} and \weak{} systems to be ${\cal K}_b =0.38$ and $0.56$ N/m, respectively. It should be noted that the computed values of ${\cal K}_b$ are similar to that for the SB system which was estimated to be 1 N/m using AFM force spectroscopy~\cite{Hanley:2003iz,Florin:1994el,Liu:2010em}.

Our multiscale coupling method using the Bell potential, in which we parameterize the receptor-ligand interaction energy (eqn.~\eqref{eqn:Hbond}) directly using all atom molecular dynamics simulations, is a straightforward approach to couple the molecular scale information to the mesoscale. {In more complex scenarios where the free energy landscape cannot be approximated using simple analytical functions, such as for receptor-ligand interactions that display a catch-bond like behavior, the PMF can be directly used as a lookup table in place of Eqn.~\eqref{eqn:Hbond}. Methods to define reaction coordinates and estimate free energy landscapes for such complex systems has been previously demonstrated for a number of biomolecules, see Sotomayor and Schulten~\cite{Sotomayor:2007if} for a review.} The effective spring constant ${\cal K}_b$ may also be estimated by analyzing the equilibrium trajectories of the bound state of the receptor-ligand, as we discuss below. 

\begin{figure}[!h]
\centering
\includegraphics[width=7.5cm,clip]{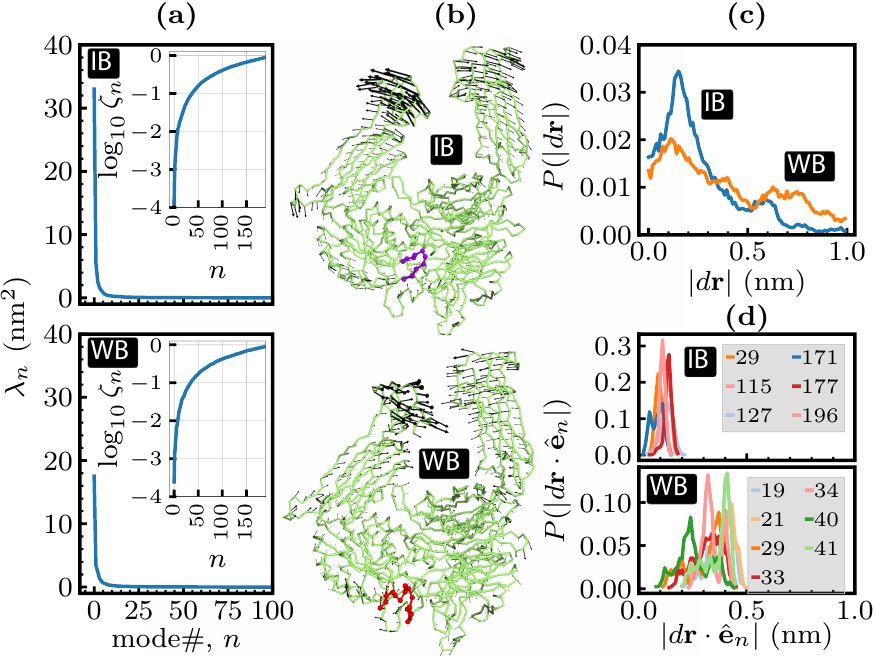}
\caption{\label{fig:pca-pot} Principal component analysis (PCA) of 10 ns equilibrium trajectories of the IB and WB systems. $\lambda_n$, for $n=1\cdots 100$, the first 100 eigenvalues of the covariance matrix, are shown in panel (a) and the corresponding spring constants ${\zeta}_n = \kbtt/\lambda_n$ are shown in the insets. Panel (b) shows the average positions of the C$_\alpha$ atoms in the IB and WB systems, and the arrows denote the direction of the first eigenvector $\hat{{\bf e}}_1$. The distribution of the average displacement ($|d{\bf r}|$) of the C$_\alpha$ atoms in the ligand are shown in panel (c), while their projections ($|d{\bf r} \cdot \hat{{\bf e}}_n|$) along various eigendirections $\hat{{\bf e}}_n$ are shown in panel (d).}
\end{figure}

\noindent\textbf{Estimating the stiffness of receptor-ligand bonds through principal component analysis (PCA):} We used PCA to analyze 10 ns equilibrium trajectories of the IB and WB systems. For each system the $3N\times 3N$ covariance matrix of atomic positional fluctuations was constructed from the positions of the $N$ $\alpha$-carbon atoms (C$_\alpha$) in the backbone of the receptor-ligand complex. Before constructing the covariance matrix, the overall translations and rotations associated with the center of mass of the protein were removed by reordering the protein backbone in all successive frames with the first frame. This results in $6$ of the $3N$ eigenvalues to be practically zero.
The $3N-6$ eigenvalues ($\lambda_1>\cdots>\lambda_{3N-6}$) and the $3N-6$ eigendirections ($\hat{{\bf e}}_1,\cdots,\hat{{\bf e}}_{3N-6}$) of the covariance matrix denote the magnitudes and directions of the $3N-6$ collective modes that dominate the equilibrium fluctuations of the bound receptor-ligand pair. Treating the system as a collection of $3N-6$ independent harmonic oscillators,  the effective stiffness of the $n^{\rm th}$  mode can be computed as $\zeta_n=\kbtt/\lambda_n$~\cite{Amadei:1993kx}. In Fig.~\ref{fig:pca-pot}(b) the ligand backbone is shown as a bead-stick representation while the receptor backbone is shown as a ribbon. In our calculations we take all  receptor C$_\alpha$ atoms within $6$ nm  of any ligand C$_\alpha$ atom to constitute the binding domain of the receptor.  Let $r$ denote the distance between the centers of masses of the receptor binding domain and that of the ligand. Our aim here is to identify which of the eigenmode(s) project significantly along r.

$\lambda_n$ for the first 100 principal modes for the IB and WB systems are shown in the main panels of Fig.~\ref{fig:pca-pot}(a), while the corresponding values of $\zeta_n$ are shown as insets to Fig.~\ref{fig:pca-pot}(a)---the value of $\zeta_n$ for $n<200$ is in the range $10^{-4}$--$10^{0}$ N/m. In both the cases, it may be seen that $\lambda_n$ is significant only for the first 10 modes indicating that the principal dynamics are contained within these modes. Our analysis of the motion of the IB and WB systems along the first 10 principal modes shows that these modes do not significantly contribute to fluctuations in $r$. For instance, the arrows in Fig.~\ref{fig:pca-pot}(b) shows the magnitude and the direction of the first principal mode which, as may be seen for both the IB and WB systems, only alters the configuration of the receptor far from the binding site. 

We computed the measure $|d{\bf r}|$ to determine the principal modes that contribute the most to the observed fluctuations in $r$. It is defined as the displacement of the individual C$_\alpha$ atoms in the ligand with respect to the center of mass of the binding domain in the receptor. Fig.~\ref{fig:pca-pot}(c) shows $P(|d{\bf r}|)$, the distribution of $|d{\bf r}|$, for both the IB and WB systems. $P(|d{\bf r}|)$ for the WB shows a wider distribution compared to the IB and this is attributable to the weaker free energy landscape of the WB system, as was shown in Fig.~\ref{fig:molecular-pmf}. Next, the measure $|d{\bf r}|$ was projected on to each eigenvector as $d{\bf r}\cdot \hat{{\bf e}}_n $ and the distributions of the top 6 projections for the IB and WB systems are shown in Fig.~\ref{fig:pca-pot}(d). For the IB, we find the maximum displacement of the binding region to be in the range $0.1$--$0.2$ nm and distributed along principal modes with $29 \leq n \leq 196$. Similarly, for the WB we find the maximum displacements to be in the range $0.3$--$0.5$ nm and distributed between modes $19\leq n \leq 41$. From the values of the force constant ${\zeta}_n$ shown in the inset to Fig.~\ref{fig:pca-pot}(a) we estimate the effective stiffness of the receptor-ligand bond in the IB and WB systems to be in the range  ${\cal K}_b=10^{-1}$--$10^{0}$ N/m. Hence, the values of ${\cal K}_b$ estimated using PCA agrees very well with the values estimated previously using Steered MD calculations.

\subsection{Effect of substrate compliance}
The local mechanical micro-environment, which denotes the compliance of the substrate on which the target receptor molecule is expressed, is also an important factor that influences NC binding. For example, a cell membrane, a PDMA substrate and a glass surface are examples of substrates with high, nominal and zero compliance, respectively. This can further be quantified by expressing the inverse of the substrate compliance in terms of the substrate's bending rigidity $\kappa =Yh^3/12$, where $Y$ is the Young's modulus and $h$ is the substrate thickness. On this scale, receptors in their natural environment (i.e., on a cell membrane) feel a substrate with bending rigidities in the range $20$--$160$ \kbt{}, while those immobilized on a glass substrate feel a bending rigidity of $\sim 10^{18}$ \kbt{}. We have previously developed an explicit model to study the effect of membrane compliance on nanocarrier binding~\cite{Ramakrishnan:2016fl}. In this work, we account for the effect of the substrate compliance on NC binding using a mean field approach, which is computationally much more tractable than the rigorous approach in ~\cite{Ramakrishnan:2016fl}. While this implementation is new, it was inspired by a similar discussion by  Mogilner and Oster~\cite{Mogilner:1996ip}. We outline this approach next.

Consider two neighboring receptor molecules at a distance 2$x$ apart on a substrate. In our calculations, we only account for the bending modes of the substrate and neglect any contributions due to stretching --- this is a very good approximation even for highly compliant substrates such as cell membranes. The maximum radius of curvature that can be induced by independent deformations of the receptor molecules is $x$, for which the membrane bending energy is $4\pi\kappa$ assuming that the deformation is hemispherical in shape. A harmonic approximation to model this deformation can be constructed by satisfying the condition $4\pi\kappa={\cal K}_m  x^2 /2$, where $x$ is the maximum extent of deformation normal to the planar membrane. This yield the spring constant for membrane compliance as ${\cal K}_m=8\pi\kappa/x^2$. Setting $x\sim 5$--$10$ nm, equal to the average distance of separation between the receptors on the surface, we estimate ${\cal K}_m$ to be in the range $10^{-2}$--$10^{-3}$ N/m. In effect, we are accounting for the energy of substrate deformation as an independent additional contribution resulting from receptor-ligand interactions. In this mean field approach, we model a receptor-ligand bond as two springs in series --- the first spring corresponds to the explicit interaction between the receptor and ligand with a spring constant ${\cal K}_b$ and the second spring corresponds to the fluctuations in the substrate with a spring constant ${\cal K}_m$. The effective spring constant, that includes the effect of  substrate compliance, is then given by ${\cal K}_b^{\rm eff}={{\cal K}_b{\cal K}_m}/({\cal K}_b+{\cal K}_m)$, and its value in our calculations is in the range $10^{-3}$--$10^0$ N/m. In the following, we represent the interactions between a receptor and ligand in the mesoscale model using a modified form of eqn.~\eqref{eqn:Hbond} given by: \begin{equation}
 {\cal H}_{\rm bond}= \left\{  
 \begin{array}{cc}
 0 & \textrm{when } d \geq d^{*} \\
 \Delta {\cal H}_{b} + \dfrac{{\cal K}_{b}^{\rm eff}}{2} d^{2}  &  \textrm{when } d<d^{*}
 \end{array}
 \right.,
 \label{eqn:effHbond}
 \end{equation}
with $d^*=\sqrt{2\Delta{\cal H}_b/{\cal K}_b^{\rm eff}}$.
\subsection{Results from the mesoscale simulations}
We first present a brief overview of the statistical mechanics framework to evaluate the binding avidity of a functionalized NC -- see~\cite{Liu:2010em,Ramakrishnan:2016fl} for the complete derivation. The equilibrium association constant for a bound NC is given by:
\begin{equation}
K_a=\dfrac{1}{{\rm [L]}}\dfrac{{\cal P}_b}{{\cal P}_u}, 
\label{eqn:assoc}
\end{equation}
where ${\rm {[L]}}$ is the ligand concentration. The probability for one NC to be in an unbound state is given by:
\begin{equation}
{\cal P}_u =\dfrac{1}{{\cal Z}} \prod\limits_{i=1}^{N_r}\left(\int\limits_{u} d{\bf X}_{r,i} \int\limits_{u} d{\bf \Theta}_{r,i} {\rm e}^{-\beta{\cal H}_{\rm flex}}  \right) \int\limits_{u} d{\bf X}_{p} \int\limits_{u} d{\bf \Theta}_{p},  
\label{eqn:punbound}
\end{equation}
where $u$ implies that the integral is carried over all configurational degrees of freedom in which the NC is in an unbound state. ${\cal Z}$ is total partition function that accounts for both the unbound and bound states. The probability for the NC to be in a bound state with exactly $m$ multivalent bonds is given by:
\begin{eqnarray}
{\cal P}_b & =\dfrac{1}{{\cal Z}} &  \prod\limits_{i=m+1}^{N_r-m}\left(\int\limits_{u} d{\bf X}_{r,i} \int\limits_{u} d{\bf \Theta}_{r,i} {\rm e}^{-\beta{\cal H}_{\rm flex}}  \right) \nonumber \\
& &\prod\limits_{j=1}^{m}\left(\int\limits_{b} d{\bf X}_{r,j} \int\limits_{b} d{\bf \Theta}_{r,j} {\rm e}^{-\beta{\cal H}_{\rm flex}} \right)  \nonumber \\
& &\int\limits_{b} d{\bf X}_{p} \int\limits_{b} d{\bf \Theta}_{p} \exp(-\beta {\cal W}_{\rm NC}).
\label{eqn:pbound}
\end{eqnarray}
Here $b$ denotes the region of the configurational space corresponding to the bound state of the NC and ${\cal W}_{\rm NC}$ is the free energy of binding, computed using umbrella sampling and WHAM as described in the methods section. Using eqns.~\eqref{eqn:punbound} and ~\eqref{eqn:pbound} in eqn.~\eqref{eqn:assoc}, and taking ${\rm [L]}=(\int\limits_{u} d{\bf X}_p)^{-1}$, the association constant may be computed as~\cite{Liu:2010em}:
\begin{eqnarray}
 K_a & = &  
 \left(\dfrac{N_r\,!}{(N_r-\eavm{})\,!\,\,\eavm{}\,!}\right)
 \left(\dfrac{N_l}{\eavm{}}\right) \nonumber \\
 & & \nonumber \\
 & & {\left(  {\rm T1} \right)^{\eavm{}}} 
{\left( {\rm T2}  \right)^{\eavm{}}} 
\left( {\rm T3} \right)
{\left({\rm T4}  \right)}.
\label{eqn:kaexp}
\end{eqnarray}
Here, we take \avm{} as the average multivalency in the lowest free energy state and the first term denotes a binomial distribution to select \avm{} receptors out of $N_r$.  The association constant, in addition to the first two terms which arise from combinatorial entropy,  is a product of  four additional terms namely:
\begin{enumerate}[(i)]
\item ${\rm T1}={\int\limits_b dx_1dy_1}/{\int\limits_u dx_1dy_1},$ the translational entropy of bound and unbound receptors,
\item ${\rm T2}={\int\limits_b \sin\theta_1 d\theta_1{\rm e}^{-\beta{\cal H}_{\rm flex}}}/{\int\limits_u  \sin\theta_1 d\theta_1 {\rm e}^{-\beta{\cal H}_{\rm flex}}}$, the flexural entropy of bound and unbound receptors,
\item ${\rm T3}={\sigma_{\phi_p} \sigma_{\cos{\theta}_p} \sigma_{\psi_p}}/{8\pi^2}$, the rotational entropy of bound and unbound NC and 
\item${\rm T4}=\int\limits_{b}\left(dX_p dY_p\right)\int\limits_{b} dZ_p\,{\rm e}^{-\beta{\cal W}_{\rm NC}}$, the enthalpic contribution due to energy gain upon binding. 
\end{enumerate}

 These four terms can directly be computed from our mesoscale simulations and the various symbols represent the different configurational degrees of freedom, defined in the methods section.  $\sigma_{\phi_p}$ and $\sigma_{\psi_p}$  are the rms fluctuations in the Euler angles $\phi_p$ and $\psi_p$  and $\sigma_{\cos{\theta}_p}$ is the rms fluctuations in the cosine of ${\theta}_p$~\cite{Carlsson:2005jy}. $K_a$ is determined by both the entropic loss and enthalpic gain due to binding, and it should also be noted that the entropic contributions start to dominate as the multivalency \avm{} increases. In computing T$1$, we take the unbound area for the receptor $\int \limits_{u} dx_1 dy_1=L_s^2/N_r$, i.e., equal to the average area per receptor~\cite{Liu:2010em,Ramakrishnan:2016fl}. 
In the following, we will show how the multivalency, the free energy of binding and the various entropic terms (T$1$--T$3$) vary with respect to variations in the system parameters.

\begin{figure*}[!h]
	\centering
	\includegraphics[width=15cm,clip]{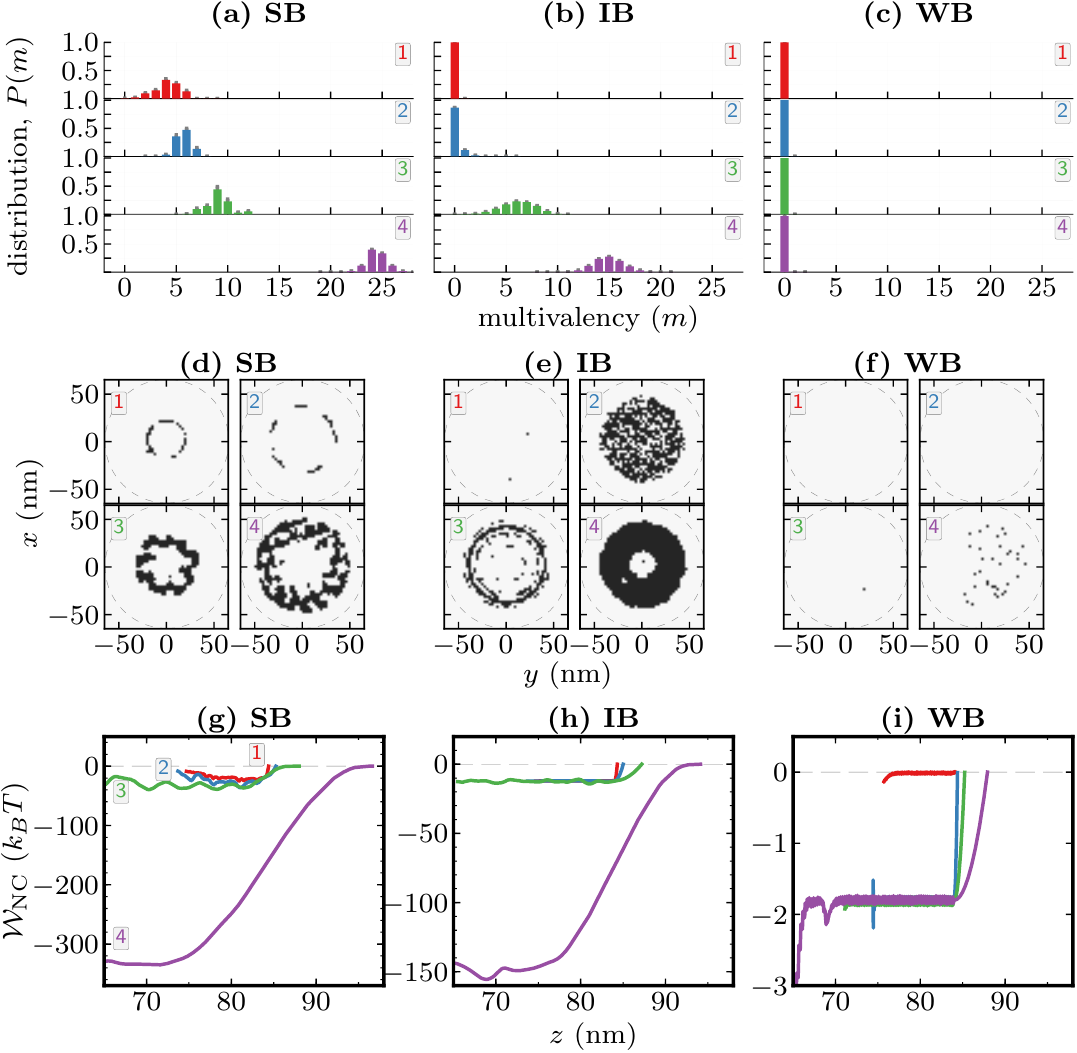}
	\caption{\label{fig:PMF-100nm} Equilibrium statistics for a 100 nm NC bound to a substrate functionalized with SB , IB  and WB systems. The four sets of data in each panel, marked \framebox{\textsf{1}}, \framebox{\textsf{2}}, \framebox{\textsf{3}} and \framebox{\textsf{4}} correspond to ${\cal K}_b^{\rm eff}=1$, $10^{-1}$, $10^{-2}$ and $10^{-3}$ N/m, respectively. Panels (a)-(c) show $P(m)$, the probability density of the NC multivalency $m$. Panels (d)-(f) show the localization density for the $x$ and $y$ positions of the bound receptors measured with respect to the center of mass of the NC, and the NC boundary is marked as solid lines. Panels (g)-(i) show the potential of mean force (PMF) as a function of the $z$ position of the NC. The PMFs are truncated at $65$ nm which is the distance of minimum approach between the NC and the substrate.}
\end{figure*}

\noindent\textbf{Effect of binding free energy and spring constant on multivalent binding of spherical NCs:}
We compare the equilibrium profiles for the SB, IB and WB systems for a NC with $a=b=c=50$ nm binding to a functionalized flat substrate. We chose the surface density of the functionalized ligands to be $14\%$, which fixes the number of ligands on the NC to be $\nab{}=162$. We first study how the energy barrier $\Delta {\cal H}_b$ and the effective spring constant ${\cal K}_b^{\rm eff}$ influence NC binding. The multivalency distribution $P(m)$, the localization profile of the bound receptors and the free energy of binding ${\cal W}_{\rm NC}$ for three different values of $\Delta {\cal H}_b$ (chosen to represent the SB, IB and WB systems) and four different values of ${\cal K}_b^{\rm eff}$ (chosen to represent the range of spring constants determined above) are shown in Fig.~\ref{fig:PMF-100nm}. The left, middle and right panels  correspond to the SB, IB and WB systems, respectively, while curves marked \framebox{\textsf{1}}, \framebox{\textsf{2}}, \framebox{\textsf{3}} and \framebox{\textsf{4}} correspond to ${\cal K}_b^{\rm eff}=1$, $10^{-1}$, $10^{-2}$ and $10^{-3}$ N/m, respectively.

$P(m)$, the probability distribution for the system to have $m$ multivalent bonds, shows a strong dependence on $\Delta {\cal H}_b$  and   ${\cal K}_b^{\rm eff}$, as may be seen in Figs.~\ref{fig:PMF-100nm}(a), (b) and (c). For a given value of ${\cal K}_b^{\rm eff}$, as expected, the SB system shows a higher multivalent binding. For all the three systems the number of multivalent bonds increases as the stiffness of the receptor ligand bonds decreases: e.g. in Fig.~\ref{fig:PMF-100nm}(a), we observe a peak multivalency of $m\sim 5$ for ${\cal K}_b^{\rm eff}=1$ N/m  which increases to $m\sim 25$ for ${\cal K}_b^{\rm eff}=10^{-3}$ N/m. We observe the SB system to always be in a bound state for all values of ${\cal K}_b^{\rm eff}$; this is characterized by $P(m=0)=0.0$ for all  panels in Fig.~\ref{fig:PMF-100nm}(a), while the binding statistics for the IB and WB systems are strongly dependent on ${\cal K}_b^{\rm eff}$. The IB system is observed to be fully bound only when ${\cal K}_b^{\rm eff} \leq 10^{-2}$ N/m (see Fig.~\ref{fig:PMF-100nm}(b)), while the WB system remains primarily in an unbound state for all values of ${\cal K}_b^{\rm eff}$ as shown in  Fig.~\ref{fig:PMF-100nm}(c). These results demonstrate the fact the binding behavior of a functionalized NC  depends on both the strength and stiffness of the receptor-ligand bond.

The dark spots in the colormaps shown in Figs.~\ref{fig:PMF-100nm}(d), (e) and (f) denote the positional projection to the $xy$-plane of the tip of the bound receptors with respect to that of the center of mass of the NC. The projection of the NC boundary including the size of the functionalized ligands are shown as solid circles. For all  systems studied here, the localization pattern of the bound receptors is influenced by both $\Delta {\cal H}_b$ and  ${\cal K}_b^{\rm eff}$. Ligands bound to NCs with higher multivalency, (e.g. the SB in  Fig.~\ref{fig:PMF-100nm}(d)), display \textit{annulus-like patterns} whose density increases with increasing multivalency. In contrast, NCs that bind with lower multivalency, (e.g. the top panels in Fig.~\ref{fig:PMF-100nm}(e) and all panels in Fig.~\ref{fig:PMF-100nm}(f)), show \textit{diffuse-circular patterns}. The circular- and annulus-like localization profiles for a spherical NC denote two distinct states of NC binding; the  former is mediated primarily by ligands in the vicinity of the poles while the latter is mediated by ligands in the vicinity of the equator. Using the method outlined previously~\cite{Liu:2010em,Ramakrishnan:2016fl} we will later show how such spatial patterns can be used to evaluate the loss of translational entropy of bound receptors on the substrate. It should also be noted that the \textit{annulus-like pattern} had previously been observed in the synapse of T-cells with their receptors~\cite{Qi:2001cf}.

Next we analyze the energy landscape for NC binding by explicitly computing the potential of mean force (PMF) \molnc{} as described in the methods section,  also see~\cite{Liu:2010em,Ramakrishnan:2016fl}.  \molnc{} is a measure of the enthalpic contribution to binding and it is related to the binding efficacy through the T$4$ term in Eqn.~\eqref{eqn:kaexp}. For systems with low multivalency (where the T$1$, T$2$ and T$3\, \sim 1.0$), \molnc{} may be used as a direct readout for the efficacy of the NC multivalent binding. The PMF profile for the SB, IB and WB systems are shown in Figs.~\ref{fig:PMF-100nm}(g), (h) and (i), respectively. It should be noted that in our model \molnc{} is non-zero only when the average multivalency $\langle m \rangle >0$. This feature is explicitly seen for the case of ${\cal K}_b^{\rm eff}$=1 N/m in  Fig.~\ref{fig:PMF-100nm}(i) where $\molncc{}(z)=0$ for all values of $z$.

For the SB system, displayed in Fig.~\ref{fig:PMF-100nm}(g), \molnc{} shows a very strong dependence on ${\cal K}_b^{\rm eff}$. A comparison of the PMF profiles for the SB system shows: (i) for stiffer springs (${\cal K}_b^{\rm eff}$=1.0 and $10^{-1}$ N/m), the binding energy landscape is highly complex and contains a number of distinct energy wells; each of these wells denote distinct multivalent states of the NC, (ii) as ${\cal K}_b^{\rm eff}$ decreases (e.g. for $10^{-2}$ and $10^{-3}$ N/m) the distinct energy wells overlap resulting in a smooth energy landscape (iii) the PMF well depth is a strong function of  ${\cal K}_b^{\rm eff}$ and shows a non-linear increase with decreasing  ${\cal K}_b^{\rm eff}$. The increase in well depth observed in our studies points to the fact that ${\cal K}_b^{\rm eff}$ can be used a tunable parameter in the design of functionalized nanocarriers; this can be modulated by attaching the ligand to the NC via tethers of different molecular weight and stiffness. The PMFs for the IB system shown in Fig.~\ref{fig:PMF-100nm}(h) also displays the various features noted for the SB system except that they do not contain any of the rugged free energy landscapes seen at higher values of ${\cal K}_b^{\rm eff}$. This may be attributed to the weak binding interactions between the NC and the receptors, as is evidenced by the nearly zero multivalency profile shown in Fig.~\ref{fig:PMF-100nm}(b). We also observe that for all values of ${\cal K}_b^{\rm eff}$, the PMF well depth for the IB system is smaller compared to that for the SB; e.g. for ${\cal K}_b^{\rm eff}=10^{-3}$ N/m, we observe a well depth of $-155$ \kbt{} and $-340$ \kbt{} for the IB and SB systems, respectively.
However these trends do not hold for the WB system shown in Fig.~\ref{fig:PMF-100nm}(h). Here, we find \molnc{} to be weakly sensitive to changes in ${\cal K}_b^{\rm eff}$ since the binding energy ($\sim -1.8$ \kbt{}) is indistinguishable from thermal noise. The distinct energy wells seen in some of the PMF curves, for the WB and SB systems, have a depth of $\sim$1 \kbt{} and hence are not statistically significant. These results indicate that when $\Delta {\cal H}_b$ is small, the effective spring constant for a receptor-ligand interaction is an insensitive parameter, one that does not benefit from tuning.

\begin{figure}[!h]
	\centering
	\includegraphics[width=7.5cm,clip]{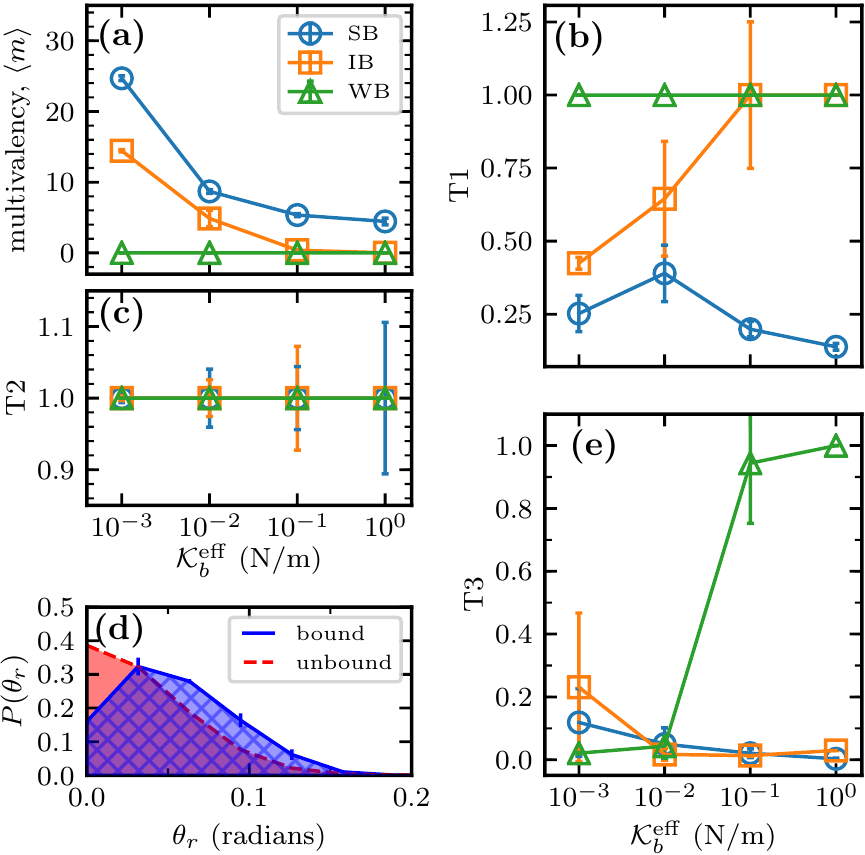}
	\caption{\label{fig:entropy} (a) The average multivalency, $\langle m \rangle$, and the various loss in entropies, (b) T$1$, (c) T$2$ and (e) T$3$, are shown as a function of the effective spring constant ${\cal K}_b^{\rm eff}$. The terms T$1$, T$2$ and T$3$ are dimensionless. Panel (d) shows the distribution of the receptor flexure angle $\theta_r$ both in the bound (solid line) and unbound (dashed line) states. The shaded regions denote the corresponding areas used to calculate T$2$.} 
\end{figure}

The enthalpic measures of the free energy \molnc{} depicted in Figs.~\ref{fig:PMF-100nm}(g)--(i) do not tell the whole story because there is significant enthalpy-entropy compensation in these systems. That is, a gain in T$4$ due to multivalent binding with large \avm{} also leads to a loss in the configurational entropies which can offset each other. We discuss the entropies next by computing the terms T$1$--T$3$ in eqn.~\eqref{eqn:kaexp} for the data shown in Fig.~\ref{fig:PMF-100nm}. The multivalency and the various losses in receptor translational, receptor flexural, and NC rotational entropies T$1$, T$2$ and T$3$  are shown in Fig.~\ref{fig:entropy}, as a function of the effective spring constant ${\cal K}_b^{\rm eff}$. The average multivalency  $\langle m \rangle$ for each of the three systems shown in Fig.~\ref{fig:entropy}(a) was directly computed from the umbrella sampling trajectories and in this calculation  we weighted the statistics of each window with its corresponding Boltzmann factor $\exp(-\beta \molncc{})$. The values of \avm{} computed using this technique agrees very well with that shown in Figs.~\ref{fig:PMF-100nm} (a), (b) and (c) which were directly computed using equilibrium simulations. This agreement establishes the consistency between the umbrella sampling and the equilibrium simulations in terms of the accessed  configurational space and establishes the ergodicity of the system under the conditions explored. As before, for the SB and IB systems we find \avm{} to  decrease with increasing ${\cal K}_b^{\rm eff}$, while for the WB system we find $\langle m \rangle \sim 0$ for all values of ${\cal K}_b^{\rm eff}$. 

\noindent\textbf{Translational entropies of the bound and unbound receptors:} The T1 term in Eqn.~\eqref{eqn:kaexp} for the SB, IB and WB systems are shown in Fig.~\ref{fig:entropy}(b). In computing T$1$, we took the unbound receptor area to be  $\int\limits_{u}dx_1 dy_1=L_s^2/N_r$, as described in our discussions on Eqn.~\eqref{eqn:kaexp},  and computed the area of the bound receptor $\int\limits_{b}dx_1 dy_1$ directly from the localization profiles shown in Figs.~\ref{fig:PMF-100nm}(d), (e) and (f); here for each system, the total area of the dark regions was taken to be $\langle m \rangle (\int\limits_{b}dx_1 dy_1)$. For the WB system,  we find T$1\sim 1.0$ for all values of ${\cal K}_b^{\rm eff}$ and this is consistent with the observation in Fig.~\ref{fig:entropy}(a), where $\langle m \rangle \sim 0$ for all cases of the WB system. On the other hand, for the SB system we observe T$1\sim 0.25$ for all values of the spring constant reflecting the fact that the higher multivalencies (Fig.~\ref{fig:entropy}(a)) and the stronger binding free energies (Fig.~\ref{fig:PMF-100nm}(g)) seen for the SB system severely constrain the accessible degrees of freedom for a bound receptor. For the IB system, we find T$1\sim 0.5$ for ${\cal K}_b^{\rm eff} \leq 10^{-2}$ N/m and T$1 \sim 0.0$ for higher spring constants and this is a signature of a transition from a bound to an unbound regime. The large error bars seen for ${\cal K}_b^{\rm eff} \leq 10^{-1}$ N/m is due to the occasional transition of the NC between its small number of bound and large number of unbound states and this is consistent with the multivalency distribution and localization profile shown in  Figs.~\ref{fig:PMF-100nm}(b) and (e), respectively. It is evident from Eqn.~\eqref{eqn:kaexp} that a reduction in the value of T$1$ also reduces the equilibrium association constant $K_a$ but the magnitude by which T$1$ influences $K_a$ varies from system to system; e.g., for the SB system with T$1=0.25$ and $\langle m \rangle =10$, the expected reduction in $K_a$ due to loss in receptor translational entropy is $(0.25)^{10} \sim 10^{-6}$, while for the IB system with T$1=0.5$ and similar multivalency the expected reduction is $(0.5)^{10} \sim 10^{-3}$. This is a general phenomenon that is valid for all configurational entropy terms and quantifies the enthalpy-entropy compensation in these multivalent systems.

\noindent\textbf{Flexural entropies of the bound and unbound receptors:} The T$2$ term was computed from the probability distribution of $\theta_r$, the polar angle of the receptor, which for the bound and unbound states are shown in Fig.~\ref{fig:entropy}(d) as solid and dashed  lines, respectively. The unbound distribution, as expected from ~\eqref{eqn:Hflex}, follows the Boltzmann statistics $\exp(-\beta{\cal H}_{\rm flex})$, with a peak value at $\theta_r=0$.  On the other hand, the peak value for the bound receptors shifts to $\theta_r \sim 0.03$ thereby constraining the  flexural degrees of freedom. The bound and unbound state integrals in the T$2$ term were taken to be the numerically computed areas under the solid and dashed curves, respectively. In Fig.~\ref{fig:entropy}(c) we display the T2 term in Eqn.~\eqref{eqn:kaexp} as a function of ${\cal K}_b^{\rm eff}$ for the SB, IB and WB systems.  We find T$2\sim 1.0$ for all the three systems which indicates that not only the flexural degrees of freedom do not incur any entropy loss upon multivalent NC binding, but also that the receptor flexural degrees of freedom are not severely impacted by changes in $\Delta{\cal H}_b$ and ${\cal K}_b^{\rm eff}$. However, there is a noticeable increase in receptor flexure for ${\cal K}_b^{\rm eff}>0.1$ which is reflected in the large error bars seen in this regime. The increase in the fluctuations of the receptor flexure is a signature of flexural contributions to multivalent binding. In other words, since the cutoff distance for bond formation $d^*$ is extremely small when ${\cal K}_b^{\rm eff}>0.1$ ($0.3$ and $1.5$ nm for the SB and IB systems) the receptors promote binding within this small cutoff distance by accessing more flexural degrees of freedom.

\noindent\textbf{Rotational entropies of the  NC:} We estimated the rotational volume ($\sigma_{\phi_p} \sigma_{\cos{\theta}_p} \sigma_{\psi_p}$) of a bound NC from the fluctuations of its Euler angles about the NC center of mass, as described in the methods section and in Eqn.~\eqref{eqn:kaexp}, also see \sinfo{S5}. The computed values of the T$3$ term are displayed in Fig.~\ref{fig:entropy}(e) for the SB, IB and WB systems. The T$3$ term clearly delineates the bound state of the NC from its unbound state, with T$3<1.0$ for the former and T$3 \sim 1.0$ for the latter. T$3$ for the WB system show a clear transition from a bound state to an unbound state at ${\cal K}_b^{\rm eff}>10^{-2}$  N/m. On the other hand, T$3$ for the SB and WB systems shows a monotonous decrease with increasing spring constant. The observed decrease in the rotational entropy  at larger values of  ${\cal K}_b^{\rm eff}$ is due to the extremely  small values of $d^*$ that constrains the rotational degrees of freedom accessible to a bound NC. T$3$ for the IB system shows an increasing trend at ${\cal K}_b^{\rm eff}>10^{0}$  N/m and this is due to constant binding and unbinding of the NC, as noted previously in our discussion regarding the multivalency under these conditions. Though the rotational entropy for a bound NC shows a large decrease from its unbound value, its contribution to $K_a$ is not as significant as that of T$1$ and T$2$ (unless the multivalency $\eavm \sim 1$), since the T$3$ term does not depend on the multivalency, see Eqn.~\eqref{eqn:kaexp}. 

\begin{figure}[!h]
	\centering
	\includegraphics[width=7.5cm,clip]{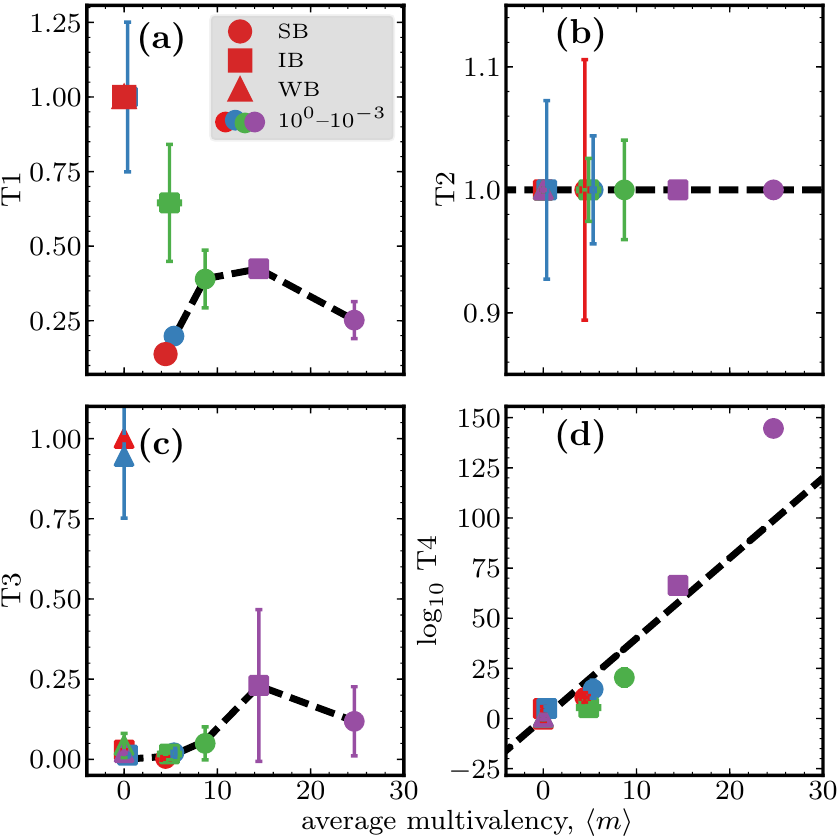}
	\caption{\label{fig:mval-integral} The configurational entropies T$1$--T$3$, shown in Fig.~\ref{fig:entropy}, and the enthalpic term T$4= \int\exp(-\beta\molncc{})dz$ as a function of the average equilibrium multivalency \avm{}. The log values of T4 are displayed and these were computed using  PMF profiles shown in Figs.~\ref{fig:PMF-100nm}(g), (h) and (i). In all panels, circle, square and triangle represent SB, IB and WB systems, respectively, while the various colors denote different values of ${\cal K}_b^{\rm eff}=10^0,\,10^{-1},\,10^{-2},\,10^{-3}$ N/m.}
\end{figure}
\subsection{Scaling behavior of the configurational entropies as a function of multivalency:} \label{sec:delgkeff} In Figs.~\ref{fig:mval-integral}(a)--(c) we display the three configurational entropies T$1$--T$3$ as function of their respective average multivalency \avm{}. T$2$ in Fig.~\ref{fig:mval-integral}(b) is insensitive to changes in \avm{}, while the values of T$1$ and T$3$ follow a peaked distribution when in the bound state; here, we define a bound state to be one in which T$_i \leq 0.5$. The bound state values of T$1$ and T$3$ were fit to an exponentially modified Gaussian distribution which are as dotted lines in Figs.~\ref{fig:mval-integral}(a) and (c).  

We next quantify the T$4$ term in Eqn.~\eqref{eqn:kaexp} which represents the enthalpic contribution to binding. In Fig.~\ref{fig:mval-integral}(d), we display T$4$ as a function of the average multivalency $\langle m \rangle$, for the SB, IB and WB systems. The different values of \avm{} correspond to simulations with varying values of ${\cal K}_b^{\rm eff}$ and this correspondence was previously shown in Fig.~\ref{fig:entropy}(a). For the WB system we find $\log_{10}$T$4\sim 0$ and this is expected since $\eavm{}\sim 0$ in all our simulations of the WB system. As expected, for the SB and IB systems  $\log_{10}$T$4$ shows a linear increase with increasing \avm{} for the most part, i.e. except for SB with a low ${\cal K}_b^{\rm eff}=10^{-3}$ N/m. It is interesting to note that all values of T$4$ computed for entirely different systems appear to follow this scaling relation that only depends on \avm{}, i.e., $\ekbt{}\log_{10}{\rm T} 4= {\cal C}\eavm{}$, with a value of ${\cal C}=3.9$ \kbt{}. Later, we will show that this scaling is universal and even applies to different ligand compositions (Sec.~\ref{sec:ligandhet}) and different sized NCs (Sec.~\ref{sec:size}). We ascertain the following by comparing the magnitudes of T$4$ to the various configurational entropies T$1$, T$2$ and T$3$, see Figs.~\ref{fig:mval-integral}(a)--(d): (i) for very small and very large values of \avm{}, the enthalpic contribution T$4$ is the major determinant of the NC association constant $K_a$, and (ii) at intermediate values, e.g., for the SB with $\eavm{} \sim 10$ for which we previously calculated $({\rm T1})^{\eavm{}}\sim10^{-6}$, the magnitudes of the entropic and enthalpic terms are comparable and hence $K_a$ is jointly determined by both these contributions. The scaling relations in Eqn.~\eqref{eqn:kaexp} and the plots in Figs.~\ref{fig:mval-integral}(a)--(d) together quantify the enthalpy-entropy compensation discussed earlier.

\subsection{Effect of particle shape anisotropy on NC binding} \label{sec:anisotropy}
In the previous section, we had shown that the nature of the molecular interactions of a receptor-ligand pair can significantly influence the binding efficacy of a functionalized particle at the mesoscale. Here we investigate how the shape of the functionalized particle is a key factor that determines the binding efficacy of a functionalized nanocarrier.  In the remainder of this article, we will primarily  use the multivalency \avm{} and the T$4$ term to  qualitatively determine the changes in the efficacy of binding  as a function of the various system parameters.

We  assess the role of particle anisotropy on NC binding by analyzing its multivalency for seven different spheroidal particles with aspect ratios $\varepsilon=$$0.1$, $0.2$, $0.5$, $1.0$, $2.0$, $5.0$ and $10.0$. The different particles were chosen to have similar volumes (equal to that for a spherical particle with $a=b=c=$ $50$ nm), similar number of ligands $\nab{}=162$, and binding to a substrate with receptor density of $2000$ receptors/$\mu{\rm m}^2$; the exact dimensions used in our simulations are given in \sinfo{S3}. Since computing PMF profiles for anisotropic particles is non-trivial, we primarily rely on the statistics of its multivalency to qualitatively characterize their binding efficacies. As previously shown in Fig.~\ref{fig:mval-integral}, we associate particle shapes that have a higher multivalency to a higher binding avidity.
\begin{figure}[!h]
	\centering
	\includegraphics[width=7.5cm,clip]{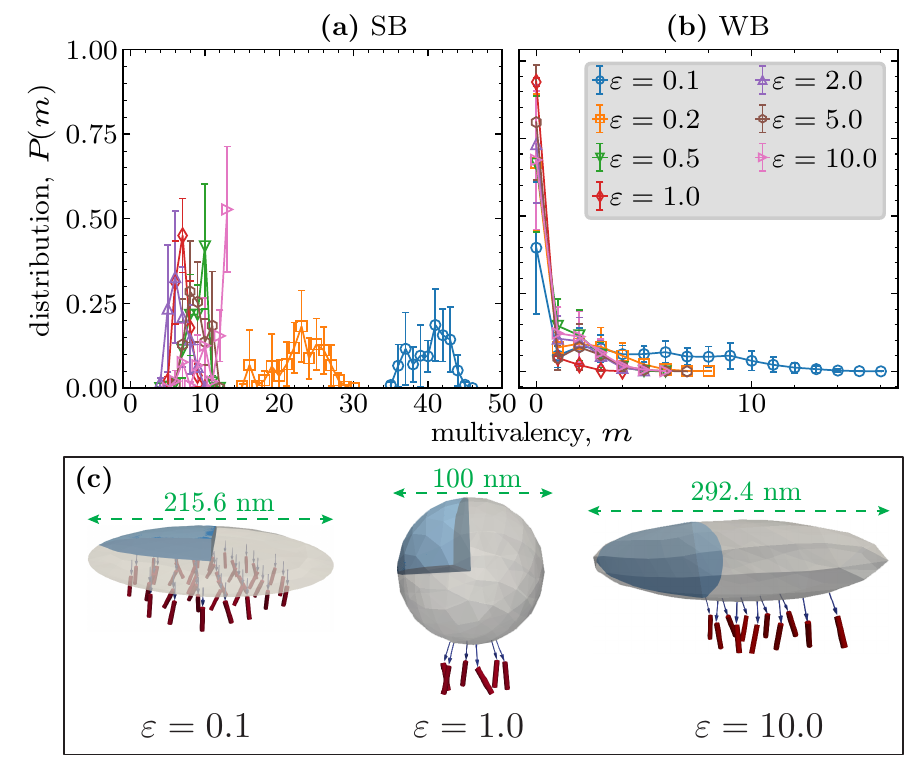}
	\caption{\label{fig:eff-aspratio} Multivalency distribution $P(m)$ as a function of the particle aspect ratio $\varepsilon$. Data shown for (a) the SB system and (b) the IB system, with $\nab{}=162$, $\nant{}=2000$ and $\varepsilon=$$0.1$, $0.2$, $0.5$, $1.0$, $2.0$, $5.0$ and $10.0$. Panel (c) shows representative snapshots of the bound state of  SB ellipsoidal NCs with $\varepsilon=$$0.1$, $1.0$ and $10.0$ --- the unbound ligands have been removed for clarity and the shaded region shows a cross section of the NC.}
\end{figure}
The multivalency distribution as a function of $\varepsilon$ for the SB and IB systems with ${\cal K}_b^{\rm eff}=0.1$ N/m is displayed in Figs.~\ref{fig:eff-aspratio}(a) and (b), respectively; additional data for ${\cal K}_b^{\rm eff}=1.0$   may be found in the \sinfo{S6}. $P(m)$ for $\varepsilon=1.0$ (shown earlier in Fig.~\ref{fig:PMF-100nm}(a) and (b)) is used as the reference to assess the effect of particle shape anisotropy. For the SB, see Fig.~\ref{fig:eff-aspratio}(a), we find the  multivalency of ellipsoidal particles, both $\varepsilon<1.0$ and $\varepsilon>1.0$, to be larger compared to that for $\varepsilon=1.0$. The oblate spheroids specifically show a higher efficacy for binding and this is exemplified in Fig.~\ref{fig:eff-aspratio}(a) for  $\varepsilon=0.2$ and $0.1$.  The role of shape anisotropy is more pronounced for the IB system shown in Fig.~\ref{fig:eff-aspratio}(b). Here, the particle is primarily in an unbound state for aspect ratios $0.2\leq\varepsilon \leq 5.0$. This may be inferred from the dominant contribution of $P(m=0)$ for these values of $\varepsilon$. A further change in the particle aspect ratio, either to more oblate, (see Fig.~\ref{fig:eff-aspratio}(a) for $\varepsilon=0.1$), or to more prolate, (see Fig.~\ref{fig:eff-aspratio}(a) for $\varepsilon=10.0$), increases the avidity of NC binding by stabilizing non-zero multivalent interactions. Our results clearly show that the shape of the ligand-coated-particle can be a key factor that determines NC efficacy, particularly when the functionalized ligand has a weak affinity for the target receptor. We also find that the oblate spheroids display a higher efficacy for binding compared to spherical and prolate-spheroidal particles. 
\subsection{Effect of ligand heterogeneity} \label{sec:ligandhet}
The binding efficacy of a functionalized NC also depends on the composition of ligands.  This is a common scenario even when the NC is functionalized with only one type of ligand, usually SB, because of the phenomenon of  opsonization~\cite{Blanco:2015ek,Moghimi:2012hy,Lundqvist:2008ge}. Here, proteins or {epitope-fragments} from the serum or plasma can deposit onto the NC surface to represent a second class of ligands that are typically IB or WB.  We assessed the role of ligand heterogeneity on  binding efficacy by studying the binding of a spherical NC, with $a=b=c=50$ nm functionalized with two types of ligands, chosen to be the SB and IB systems. The ligands were randomly distributed on the particle surface and both types of  ligands were taken to bind to the same surface receptor with corresponding values of $\Delta {\cal H}_b$, as in Table~\ref{table:system}, and with a spring constant ${\cal K} _b^{\rm eff}=1.0$ N/m. All our results correspond to the average behavior determined from four independent ensembles. We systematically varied the composition of SB:IB ligands, holding constant the total number of ligands  $N_l=162$. 
\begin{figure}[!h]
\centering
\includegraphics[width=7.5cm,clip]{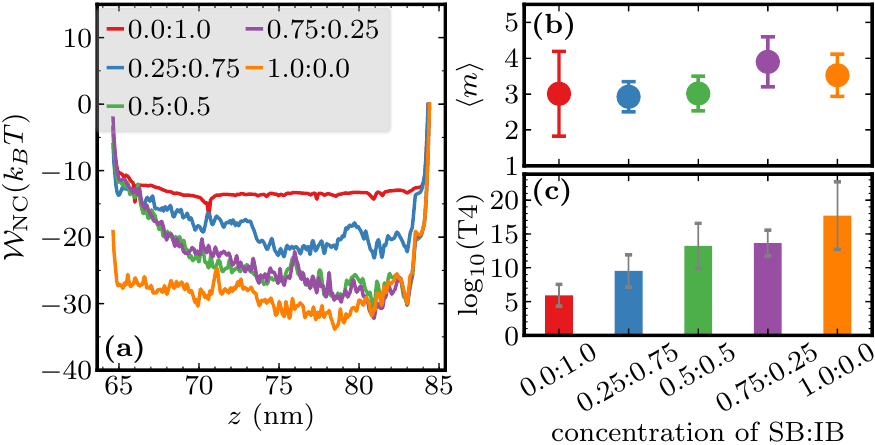}
\caption{\label{fig:eff-heterogeneity} The potential of mean force for a spherical nanocarrier, with $a=b=c=50$ nm, functionalized with varying concentrations of the SB and IB receptors. The average multivalency \avm{} and the T$4$ term computed from the PMF are shown in panels (b) and (c), respectively. Data shown for a NC with $a=b=c=50$ nm with ${\cal K}_b^{\rm eff}=1.0$ N/m.}
\end{figure}
In Fig.~\ref{fig:eff-heterogeneity}, we show the computed values of the PMF, ${\cal W}_{\rm NC}$, the average multivalency, \avm{},   and the enthalpic contribution to binding, T$4$. The corresponding data are displayed in panels (a), (b) and (c) respectively. \molnc{} is highly sensitive to changes in the ligand composition and we find its well depth increases with increasing composition of SB ligands, as shown in Fig.~\ref{fig:eff-heterogeneity}(a). The observed sensitivity is also seen in the T$4$ term shown in  Fig.~\ref{fig:eff-heterogeneity}(c) whose values are in the range $10^5$ for the fully IB system (0.0:1.0) to $10^{15}$ for the fully SB system (1.0:0.0). The corresponding changes in the T$1$, T$2$ and T$3$ terms are not significant and this is shown in \sinfo{S7}. The observed ten orders of magnitude increase in T$4$ is achieved with no considerable changes in the average multivalency, which as shown in  Fig.~\ref{fig:eff-heterogeneity}(b) is  $\eavm{}\sim 3$ for the five ligand compositions investigated here. Moreover, the values of T$4$ in Fig.~\ref{fig:eff-heterogeneity}(c) show a gradual increase in response to changes in the ligand composition, unlike those in Fig.~\ref{fig:mval-integral} that show a highly non-linear increase in response to changes in $\Delta {\cal H}_b$ and ${\cal K}_b^{\rm eff}$.  Thus, these results clearly demonstrate that the ligand composition can be used as a tunable parameter to design NCs with a wide range of binding efficacies and also can be pre-engineered to reduce the effect of opsonization that occurs \textit{in vivo}. The role of ligand heterogeneity as a control parameter for NC uptake was recently demonstrated in an experimental study by Levine and Kokkoli~\cite{Levine:2017bj}, for the case of NCs coated with ligands for $\alpha_5\beta_1$ and $\alpha_6\beta_4$ integrins.
\subsection{Effect of particle size on binding} \label{sec:size}
Specific binding through multivalent receptor-ligand interactions is effectively used by systems of varying sizes from a viral particle to a cell. The size of the particle influences a multitude of factors including changes in the number of ligands, reduction in the particle curvature, all of which can influence particle binding.  We systematically studied this phenomenon using solid spherical particles of five different sizes $a=50,\,150,\,250,\,350$ and $500$ nm, with $a=b=c$. The total number of ligands on the particles were chosen to be $N_l=162,\,1458,\,4050,\,7938$ and $16200$, respectively, such that all particles have a uniform  ligand density of $14\%$. All our calculations were performed using the SB system for two different spring constants ${\cal K}_b^{\rm eff}=0.1$ and $1.0$ N/m.
\begin{figure}[!h]
\centering
\includegraphics[width=7.5cm,clip]{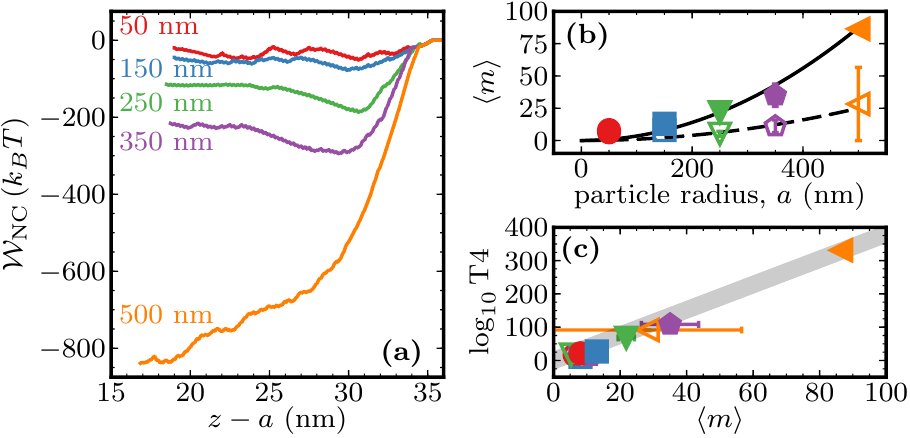}	
\caption{\label{fig:eff-size} The effect of size on the binding affinity of a functionalized particle. (a) Comparison of the potential of mean force \molnc{} for five different spherical particles with $a$=$50$, $150$, $250$, $350$ and $500$ nm. Data shown for the SB system with ${\cal K}_b^{\rm eff}=0.1$ N/m. (b) The average equilibrium multivalency \avm{} is shown as a function of $a$ while its quadratic dependence on $a$ is shown as solid and dotted lines for the two spring constants. (c) The computed values of T$4$ are plotted as a function of \avm{} and the solid band represents the relation $\ekbt{}\log_{10}{\rm T}4\sim {\cal C}\eavm{}$, with ${\cal C}\sim 3.9$ \kbt{}. The filled and open symbols in panels (b) and (c)  correspond to ${\cal K}_b^{\rm eff}=0.1$ and $1.0$ N/m, respectively.}
\end{figure}
In Fig.~\ref{fig:eff-size}(a), we show \molnc{} as a function of the scaled particle separation $z-a$ for  ${\cal K}_b^{\rm eff}=0.1$ N/m (data for ${\cal K}_b^{\rm eff}=1.0$ N/m may be found in the \sinfo{S8}).  We find the depth of \molnc{} to increase with increasing $a$ and this effect is more pronounced for ${\cal K}_b^{\rm eff}=0.1$ N/m. The well defined minimum seen at $(z-a) \sim 30$ nm for particles with $50\leq a\leq350$ nm shows that there is a significant energy barrier for the bound particle to move towards the substrate.  However, this barrier seems to be absent for the $500$ nm particle for which \molnc{} decreases as the particle moves towards the substrate. This is consistent with our observation that the equilibrium multivalency for the $500$ nm particle also shows a huge increase, as expected for a particle very close to the substrate.  The average multivalency \avm{} for ${\cal K}_b^{\rm eff}=0.1$ and $1.0$ N/m are shown in Fig.~\ref{fig:eff-size}(b) as filled and open symbols, respectively. \avm{} increases with particle size and the largest value of $\eavm \sim 80$ is seen for the $500$ nm particle with ${\cal K}_b^{\rm eff}=0.1$ N/m. The observed increase in \avm{} as a function of $a$ can be fully attributed to the corresponding increase in the total number of ligands that scales as $N_l \propto a^2$. If \avm{} increases primarily due to an increase in the number of ligands that contribute to binding we expect $\eavm \propto a^2$. The computed values of \avm{} are in excellent agreement with this relation as shown in Fig.~\ref{fig:eff-size}(b) where the dashed and solid lines are the best fit parabolic curves for  ${\cal K}_b^{\rm eff}=0.1$ and $1.0$ N/m, respectively.  

In Fig.~\ref{fig:eff-size}(c), we plot T$4$ as a function of \avm{} for different particle sizes and spring constants. The filled and open symbols correspond to data for ${\cal K}_b^{\rm eff}=0.1$ and $1.0$ N/m, respectively. Intriguingly, the data collapses onto the same scaling relation, $\ekbt{}\log_{10}{\rm T}4 = {\cal C}\eavm$ with ${\cal C}=3.9$ \kbt{}, as shown in Sec.~\ref{sec:delgkeff}. The observed scaling relation has a direct and simple interpretation, that the total binding free energy of the particle is $\eavm {\cal C}$ and the equilibrium bound state of each of the \avm{} multivalent bonds correspond to an average energy ${\cal C}$. 
\section{Conclusions}
\begin{figure}[!h]
\centering
\includegraphics[width=7.5cm,clip]{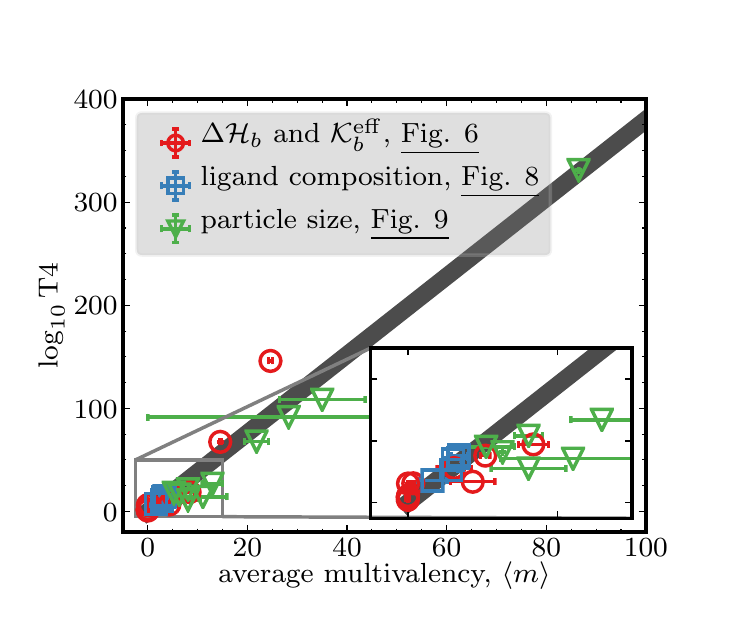}
\caption{\label{fig:mval-T4-scaling} The enthalpic contribution T$4$ for functionalized spherical particles is plotted as a function of the average multivalency \avm{}. The different sets of data were taken from Figs.~\ref{fig:mval-integral},~\ref{fig:eff-heterogeneity} and ~\ref{fig:eff-size}. The solid diagonal band represents the scaling relation  $\ekbt{}\log_{10}{\rm T}4 = {\cal C}\eavm$, with ${\cal C}\sim 3.9$ \kbt{}. {The inset shows the validity of the scaling behavior for average multivalencies in the range $0<\eavm{}<15$.}}
\end{figure}
We have presented a multiscale computational framework to study the multivalent adhesion of a rigid ligand-coated particle adhering to receptors on a substrate. In our approach, the molecular nature of receptor-ligand interactions has been retained by carrying out molecular dynamics simulations of monovalent receptor-ligand interactions. The results of the molecular dynamics simulations were directly used to parameterize the coarse-grained potentials. Our results show that the binding avidity of nano-sized functionalized carriers is primarily determined by the dominant  enthalpy-entropy compensation seen in these systems. We have demonstrated this feature by explicitly computing the free energy for multivalent binding and also the losses in configurational entropies T$1$, T$2$ and T$3$ associated with the configurational variables for receptor diffusion, receptor flexure, and rotational motion of the nanocarrier, respectively. Of these, T$1$ is the most dominant entropic term, particularly for systems with large number of multivalent interactions \avm{}, since losses in T$1$ penalizes the binding avidity as T$1^{\eavm{}}$. The effect of particle shape anisotropy on NC binding was shown to promote higher multivalency (\avm{}) for both prolate and oblate particles, compared to their spherical counterpart. Moreover, we have shown that the ligand composition, substrate compliance, tether elasticity and particle size can also be used as tunable design parameters to modulate the overall binding avidity of the NC. 

Our study suggests that each of the entropic and enthalpic terms that govern the binding avidity of the functionalized particle may follow an universal scaling behavior that only depends on the multivalency of the particle in its bound state. For instance, in Fig.~\ref{fig:mval-T4-scaling} we show this feature for the enthalpic contribution T$4$ computed for spherical particle under  completely different conditions; i.e., the three sets of data corresponding to: (i) analysis of the effect of $\Delta{\cal H}_b$ and ${\cal K}_b^{\rm eff}$ (Fig.~\ref{fig:mval-integral}), (ii) the effect of ligand composition (Fig.~\ref{fig:eff-heterogeneity}), and (iii) the effect of particle size (Fig.~\ref{fig:eff-size}). T$4$ for all these system fall on a universal master curve $\ekbt{}\log_{10}{\rm T}4 = {\cal C}\eavm$, as discussed earlier, with a single value of ${\cal C} \sim 3.9$ \kbt{}. The presence of such an universal scaling is surprising and would imply that we can describe the thermodynamics of binding merely by computing the multivalency of the particle. This is a welcome finding as it would significantly reduce the computational efforts required to estimate the binding avidity; i.e., the multivalency can simply be obtained using one equilibrium simulation rather than resorting to the umbrella sampling necessary for computing the entire free energy landscape. Whether all types of functionalized particles, independent of their shape, size and ligand heterogeneity, follow this scaling behavior and hence making it truly universal is to established and still remains an open question. A closer look at the inset to Fig.~\ref{fig:mval-T4-scaling} shows that some values of T4, particularly those from Fig.~\ref{fig:eff-size}, show deviations larger than one standard deviation (but agree to within 2 standard deviations) from the proposed scaling relation and these outliers corresponds to NCs with sizes in the range $150$ -- $350$ nm. The observed deviations are attributed to not being able to define a characteristic multivalency that represents the average behaviour because of the degeneracy in the equilibrium multivalency distributions. This is displayed in the Suppl. Info., \sinfo{S9}, where the multivalency distribution for particles that adhere to the scaling relation (for $a=50$ and $500$ nm) are unimodal, while those for particles that deviate from the scaling relation ($a=150$ -- $350$ nm) are multimodal. The presence of degenerate states leads to broad multivalency distributions and hence a deviation from the scaling relation, which is plotted using the average multivalency \avm{}.

Despite the large advances in the design of functionalized nanocarriers, the development of highly efficient and robust targeted delivery systems still remains a challenge. This is commonly attributed due to our rudimentary understanding of complex factors such as physiological conditions, cell heterogeneity, tissue architecture and target chemistry ~\cite{Mitragotri:2014by,Shi:2016fg}. Therefore, it is essential to develop quantitative models for nanocarrier dynamics and adhesion in its natural environment that can help comprehend these inherent complexities. The work presented in this article should be viewed as a step in this direction. {In our current model, while the entropy calculations are extensive, the treatment based on bell bond~\cite{Hanley:2003ga} and softening of spring due to membrane is simplistic in some scenarios like the IB system. More complex bonds like catch bonds, entropy of membrane, etc. can be added to our framework by following and incorporating models published in the literature~\cite{Peng:2012kz,Irvine:2002ix,Zhao:2009ka,Ramakrishnan:2016fl}. Atomistic simulations can be used to further refine these models and include biochemical specificity~\cite{Sotomayor:2007if,Krammer:1999cy}.} Despite the minimalistic nature of our model, it has helped us gain some key insights into the role and magnitudes of configurational entropies on NC binding which are non-trivial and non-intuitive, relative to the more intuitive (but nevertheless hard-to-quantify) multivalent enthalpic contributions. It should be noted that the effect of configurational entropies is particularly pronounced for particle dimensions in the range $50$--$350$ nm which is the preferred size for targeted carriers in most applications, making this study highly relevant for nanoparticle design. Other complexities such a physiological and hydrodynamic barriers can be added to the design in a modular fashion and integrated with the multivalent configurational entropy and enthalpy module we have presented in this work in order to develop a general and flexible computational platform for the design of functional nanocarriers for applications in nanomedicine.
\section{Authors Contributions}
M.M, A.R. R.R and N.R designed research and M.M, S.M.H and N.R performed the simulations. N.R and R.R derived all the theoretical expressions including scaling analysis and devised methods to compare to simulations including error analysis. All authors were involved in the analysis and interpretation of data and in writing of the manuscript.
\section{Acknowledgements}
This work was supported in part by Grants NSF/DMR-1120901, NSF/CBET-1236514,  NIH/U01EB016027 and 1U54CA193417. Computational resources were provided in part by the Grant MCB060006 from XSEDE. The authors declare no competing interests. We thank the anonymous referee for pointing out the subtle features in the scaling relation in Fig.~\ref{fig:mval-T4-scaling}.
\bibliographystyle{biophysj}

\end{document}